\documentclass[journal]{IEEEtran}
 \usepackage{graphicx}

\ifCLASSINFOpdf
\else
\fi

\usepackage{subfigure}
\usepackage{multirow}

\begin{document}

%
\title{Smart Grid Communications: Overview of Research Challenges, Solutions, and Standardization Activities}
%
%
%

\author{Zhong Fan, Parag Kulkarni, Sedat Gormus, Costas Efthymiou,
Georgios Kalogridis, Mahesh Sooriyabandara, Ziming Zhu,
Sangarapillai Lambotharan, and Woon Hau Chin
\thanks{Z. Fan, P. Kulkarni, S. Gormus, C. Efthymiou,
G. Kalogridis, M. Sooriyabandara and W. Chin are with Toshiba
Research Europe Limited, Telecommunications Research Laboratory, 32
Queen Square, Bristol, BS1 4ND, UK.
e-mail:zhong.fan@toshiba-trel.com.}
\thanks{Z. Zhu and S. Lambotharan are with Advanced Signal Processing Group, Department of
Electronic and Electrical Engineering, Loughborough University,
Leicestershire, LE11 3TU, UK.}}

\maketitle

\begin{abstract}
Optimization of energy consumption in future intelligent energy
networks (or Smart Grids) will be based on grid-integrated
near-real-time communications between various grid elements in
generation, transmission, distribution and loads. This paper
discusses some of the challenges and opportunities of communications
research in the areas of smart grid and smart metering. In
particular, we focus on some of the key communications challenges
for realizing interoperable and future-proof smart grid/metering
networks, smart grid security and privacy, and how some of the
existing networking technologies can be applied to energy
management. Finally, we also discuss the coordinated standardization
efforts in Europe to harmonize communications standards and
protocols.
\end{abstract}


%
\IEEEpeerreviewmaketitle

\section{Introduction}

\IEEEPARstart{C}{limate} change and greenhouse gas emissions have
become a recognized problem of international significance in recent
years.  Renewable energy sources offer a key solution to this
problem; however, their integration into existing grids comes with a
whole new set of barriers, such as the intermittency of generation,
the high level of distribution of the sources and the lack of proven
distributed control algorithms to manage such a highly distributed
generation base.

Historically, the electrical grid has been a {\it broadcast} grid
(i.e. few-to-many distribution), where a few central power
generators (i.e. power stations) provide all the electricity
production in a country or region, and then `broadcast' this
electricity to the consumers via a large network of cables and
transformers. Based on load forecasting models developed over time,
the utility providers generally over-provision for the demand
(considering peak load conditions). If the demand increases above
the average, they may have to turn on the peaker
plants\footnote{Peaker plant is a standard term used in the power
grid community. Peaker plants are switched on to meet a shortfall in
supply, on a timescale varying from a few seconds up to a few
minutes \cite{parc} (cf. Table 1 in this document).} which use
non-renewable sources of energy (e.g. coal fired plants) to generate
additional supply of energy to cope with the demand. The
provisioning for peak load approach is wasteful when the average
demand is much lower than the peak because electricity, once
produced, has to be consumed as grid energy storage is expensive
\cite{lindley}. Secondly, setting up and maintaining the peaker
plants is not only environmentally unfriendly but also expensive.
Also, given the increasing demand for energy, it may be difficult,
perhaps impossible in the longer run, to match the supply to this
peak demand. What is attractive in such a situation then, is to
match the demand to the available supply by using communication
technology (two way communications between the grid and the customer
premises) and providing incentives (e.g. through variable pricing)
to the consumer to defer (reschedule) the load during times when the
expected demand is lower so as to improve utilization of the
available capacity. This necessitates the flow of metering
information from the customer premises to the grid to identify the
demand, and control information (e.g. pricing information) in the
other direction to coerce the customer into adapting their demand.
As mentioned earlier, since the legacy grid is a broadcast grid,
this motivates the need for a communications infrastructure and
protocols to support the aforementioned functionalities.

While the legacy grid has served well for the last century or so,
there is a growing need to update it, from the points of view of
both the aging infrastructure and the new environmental and societal
challenges. As a result, national governments and relevant
stakeholders are making significant efforts in the development of
future electrical grids or Smart Grids. A smart grid is an
intelligent electricity network that integrates the actions of all
users connected to it and makes use of advanced information,
control, and communications technologies to save energy, reduce cost
and increase reliability and transparency. Development of this new
grid will require significant efforts in technology development,
standards, policy and regulatory activities due to its inherent
complexity.

A proper demand management through the smart grid technology has the
potential to yield significant savings in the generation and
transmission of energy. This is mainly due to the reduction of
number of peaker plants needed to cater for peak demand that occurs
only a small percentage of time. For example, it has been reported
in \cite{faruqui} that in Europe, five to eight percent of installed
capacity is used only one percent of the time. By deferring the peak
demand to off peak times, the capacity and transmission cost could
be reduced up to 67 billion euros in Europe \cite{faruqui}. An
annual potential value generation up to 130 billion dollars by 2019
has been forecast by McKinsey in \cite{booth} for a fully deployed
smart grid in the US. The work in \cite{netl} states that even a
conservative estimate of potential saving due to grid modernization
is 40 billion dollars per year. In addition to the direct savings,
there are many important economical and societal benefits such as
reduction of CO2 emissions, integration of renewable energy,
elimination of regional blackouts and reduced operational costs via
for example automated meter readings.

Many countries are currently making massive investments on smart
grid research and development. For example, the smart grid is a
vital component of President Obama's comprehensive energy plan: the
American Recovery and Investment Act includes \$11 billion in
investments to ``jump start the transformation to a bigger, better,
smarter grid''. One of the key elements behind the current intensive
work program towards smart grid in the United States is tightly
linked with the need to modernize their power system. In particular,
the lack of electricity distribution network reliability under
stress conditions was born out of under-investment in the
infrastructure combined with growing energy demand. This was
emphasized in a series of major supply disruption events (e.g. the
North-East blackout), widely seen as a wake-up call to address
network stability by increasing inter-connectivity, local and wide
area control. Also, there are growing expectations on the
integration of a wide range of renewable energy sources with the
power grid. Therefore the US DoE Smart Grid Research and Development
Program \cite{doe} has set the following performance targets for
2030: 20\% reduction in the nation's peak energy demand; 100\%
availability to serve all critical loads at all times and a range of
reliability services for other loads; 40\% improvement in system
efficiency and asset utilization to achieve a load factor of 70\%;
20\% of electricity capacity from distributed and renewable energy
sources (200 GW). There are also a number of huge industrial
research projects currently underway, for example, the IBM GridWise
project \cite{ibm} and the smart grid trial in New Mexico
\cite{mexico}.

Europe, by contrast, presents a highly interconnected, mesh
distribution network exhibiting more robustness than the US system.
Energy network development is in a period of stability within the
European Union (EU). There is a program of large investments in
updating the distribution network already agreed at the EU level
\cite{ec1}, which is decreasing the pressure for rapid decisions
towards adoption of disruptive new technology. The biggest concern
in Europe is in the integration of renewable power generation to
meet the 2020 targets for reduction of CO2 emissions from fossil
power generation. The intermittent nature of these energy sources
places demands that existing transmission and distribution networks
have not been designed to meet. Considerable effort will be needed
so as to become less dependent on conventional and foreign sources
of energy. The important role of smart grids is mentioned in the
European Commission's 2020 strategy document \cite{ec2}, in the EU
Smart Grids Technology Platform \cite{ec1}, and also highlighted in
the new initiative on Future Internet research (FI PPP,
\cite{fippp}) as a key application. A recent example FI PPP project
on smart grid is FINSENY led by Siemens \cite{siemens}. The EU,
through the technology development platform, has established a
carefully planned approach to the implementation of smart grid
technologies in the medium to long term. Establishing work on
standardization, research projects involving academia with
industries (utilities and manufacturers), and demonstration/pilot
projects are the current priority.

Smart grids and smart metering are expected to contribute
significantly towards improving energy usage levels through the
following four mechanisms:
\begin{itemize}

\item Energy feedback to home users through an IHD (In-Home
Display) - Accurate energy consumption, coupled with real-time
pricing information is expected to reduce energy usage within the
home, especially as energy prices continue to rise.

\item Energy consumption information for building operators - to assist with the
detailed understanding and optimization of energy requirements in
buildings.

\item The inclusion of distributed micro-generation based on
locally-distributed clean, renewable energy sources such as wind and
solar.

\item Real-time demand response and management strategies for lowering
peak demand and overall load, through appliance control and energy
storage mechanisms (including electrical vehicles).

\end{itemize}

To enable the above functionalities, an effective, reliable, and
robust communication infrastructure has to be in place. This paper
provides an overview of the following important issues of smart grid
communications: communication infrastructure, network architecture,
demand response management, security and privacy challenges, and
standardization activities. Compared to other recent surveys on
smart grid (e.g. \cite{91} and \cite{92} which are mainly from an
academic perspective), our primary aim is to provide a coherent
picture of the current status of smart grid communications,
especially focusing on research challenges, standardization, and
industry perspectives. We would like to point out that since the
smart grid is a vast area, the main focus of this paper is on smart
grid communications. For overviews on other aspects of smart grid,
e.g. technologies on the transmission side and control center of the
smart grid, please refer to \cite{13} and \cite{14}. Furthermore,
this article mainly provides a technical perspective of the smart
grid. For a business or economic perspective, the readers are
referred to \cite{15}.

The rest of the paper is organized as follows. Section 2 discusses
several research challenges and opportunities in smart grid
communications. Section 3 addresses the important issue of security
and privacy and Section 4 presents our vision of applying some
existing networking technologies to solve energy management
problems. Section 5 provides a brief introduction to the
standardization activities in Europe and conclusions are drawn in
Section 6.

\section{Communication challenges and issues}

While communications technology is seen as an essential enabling
component of future smart grids, there are a number of challenges
that must be addressed in order to have fully robust, secure and
functional smart grid networks. Some of these challenges are
discussed below.  It is important to note that these challenges are
very much intertwined, i.e. they affect each other and must be
considered as parts of a bigger problem/challenge. We begin by first
giving an overview of smart metering communications which is a major
component of the overall smart grid communications architecture.
This is then followed by a discussion on several different related
research issues \cite{16}.

\subsection{Smart metering communications}

A smart metering communication system consists of the following
components: smart meter which is a two-way communicating device that
measures energy consuming at the appliances (electricity, gas, water
or heat); Home Area Network (HAN) which is an information and
communication network formed by appliances and devices within a home
to support different distributed applications (e.g. smart metering
and energy management in the consumer premises); Neighborhood Area
Network (NAN) that collects data from multiple HANs and deliver the
data to a data concentrator; Wide Area Network (WAN) which is the
data transport network that carries metering data to central control
centers; and Gateway which is the device that collects or measures
energy usage information from the HAN members (and of the home as a
whole) and transmits this data to interested parties. Table
\ref{compare} indicates the typical communication requirements and
the potential technologies that could be employed to realize the
different types of network mentioned above. For a comprehensive
survey on communication protocols for automatic meter reading
applications, please refer to \cite{90}.

\begin{table}[ht]
  \begin{center}
\begin{tabular}{ | p{1.5cm} | p{1.5cm} | p{2cm} | p{2cm} |} \hline\hline
Type of Network & Range &  Data Rate Requirements & Potential
Technologies \\ \hline

HAN & Tens of meters & Application dependant but generally low bit
rate control information  &  ZigBee, Wi-Fi, Ethernet, PLC \\ \hline

NAN & Hundreds of meters & Depends on node density in the network
(e.g. 2Kbps in the case of 500 meters sending 60 byte metering data
every 2 minutes per NAN)  &   ZigBee, Wi-Fi, PLC, cellular \\ \hline

WAN & Tens of kilometers & High capability device such as a high
speed router/switch (a few hundred Mbps to a few Gbps) & Ethernet,
microwave, WiMax, 3G/LTE, fibre optic links \\ \hline
\end{tabular}
  \end{center}
  \caption{Communication requirements and capabilities of the different types of networks}
    \label{compare}
\end{table}

\begin{figure*}
\vspace{25mm}
 \centering
   \includegraphics[height=9.0in]{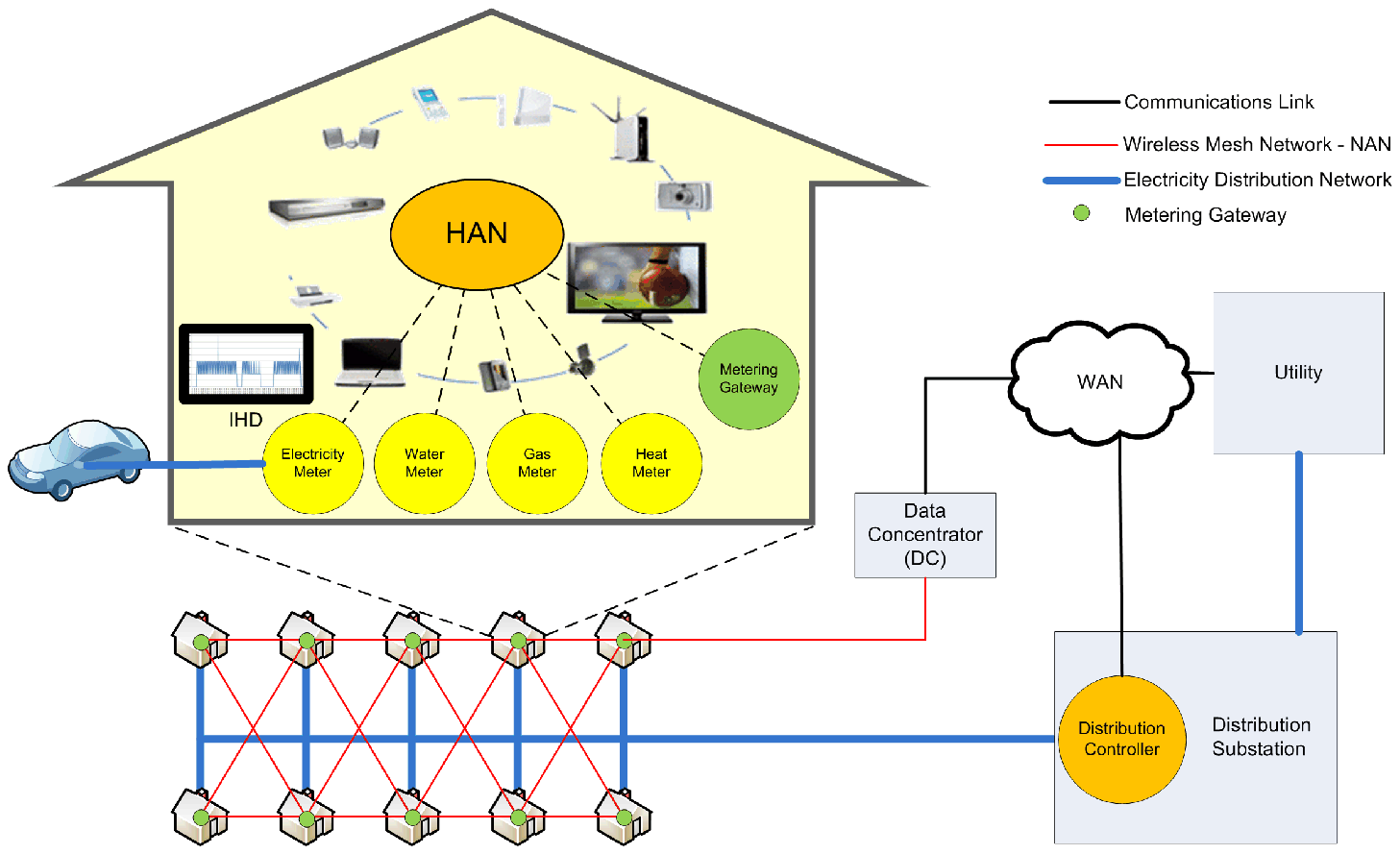}
   \vspace{-160mm}
\caption{Typical smart metering architecture} \label{sm}
\end{figure*}

Figure \ref{sm} shows a typical smart metering architecture that is
being reflected in the European standards development process.  Note
that this is just an example and not a definitive final
architecture. At the most basic level, the home will be equipped
with a series of smart meters, one each for electricity, gas, water
and heat (if applicable), according to the facilities available at
each home. These will be connected to a metering gateway in the
home, which may or may not be part of an existing home gateway
device. The HAN through which they communicate with the metering
gateway may be multi-standard. This is mainly due to differing meter
locations and power availability; for example, gas and water meters
may have to use only battery power.  Multiple HANs are further
connected into a NAN via a wireless mesh network.

In Figure \ref{sm}, the smart metering gateway is connected to both
the utility (via a WAN) and the distribution control system (DCS)
because the utility company may not necessarily own the DCS,
especially in countries such as the UK where there is so much
competition and fragmentation - a home in London could be supplied
by a Scotland-based utility with the local distribution
infrastructure being owned by another company. The utility is mainly
responsible for services like billing, service management and
tariffs, and the distribution control system is responsible for
demand response, commands to disable certain devices/appliances,
renewable energy integration, etc.

During the European standardization process, it became evident that
a single application data model is required to enhance the
interoperability of the different meters and databases used to store
their information. One such model which has been receiving a lot of
attention is the Device Language Message Specification/Companion
Specification for Energy Metering (DLMS/COSEM) model
\cite{17}\cite{18}\cite{85}, which is currently being modified to
address all of the additional functionalities that have been
mandated.

Optionally, home devices and appliances may also be part of the HAN,
whilst any home automation system(s) may be connected to the same
HAN and interface with the smart meters. The in-home display (often
called the Customer/Consumer Display Unit - CDU), which is being
considered strongly to be mandated in the UK Smart Metering
Implementation Project, is an example of this.  In the future, it is
expected that home automation systems will gather detailed energy
usage information from the smart meters and also from sub-meters
attached to specific devices and appliances, so that a number of
sensors and actuators can be brought together within the home to
optimize the energy consumption of the home as a whole.  This aspect
of smart metering and home automation is crucial to the realization
of the targets of CO2 emission reduction that the EU has set.

There are a number of options currently for the communications
outside the home, i.e. between the metering gateway and the power
distribution network, utility, operators and any other authorized
parties. The obvious candidates are wireless cellular technologies
and various home broadband solutions. However, it remains to be seen
if utilities will be willing to trust the reliability and
independence of these solutions.  It is more likely that a mixture
of a wide range of technologies will be used, including proprietary
solutions for the last-mile access to the metering gateway. Data
concentrators/aggregators have been discussed (see Figure \ref{sm}),
which could be deployed around residential areas.  Given a wireless
mesh network between metering gateways and/or smart meters, these
aggregators could gather all of the required data at periodic
intervals and then send them over to the utilities through fixed
line communications.

\subsection{Interoperability}

A key feature of smart grids is the interconnection of a potentially
large number of disparate energy distribution networks, power
generating sources and energy consumers. The components of each of
these entities will need a way of communicating that will be
independent of the physical medium used and also independent of
manufacturers and the type of devices. The communication
architecture of the future smart grid is yet to be defined. As a
result, multiple communication technologies and standards could
coexist in different parts of the system. For example, short-range
wireless such as Bluetooth or UWB could be used for the interface
between meter and end customer devices, IEEE 802.15.4 (ZigBee) and
IEEE 802.11 (Wi-Fi) could be used for smart meter interfaces in the
home and local area network, and cellular wireless (e.g. GPRS, UMTS,
or 4G technologies like 802.16m and LTE) could be used for the
interface between meters and the central system \cite{81}\cite{82}.
To this end, interoperability is essential for smart metering
devices, systems, and communications architectures supporting smart
grids. This has been emphasized by the recent EU M/441 Mandate on
smart meters \cite{19}.

It can be envisaged that in a complex system such as smart grid,
heterogeneous communication technologies are required to meet the
diverse needs of the system. Therefore, in contrast to conventional
telecommunication standardization such as IEEE 802.11n or 3GPP LTE,
the standardization of communications for smart grid means making
interfaces, messages and workflows interoperable. Instead of
focusing on or defining one particular technology, it is more
important to achieve agreement on usage and interpretation of
interfaces and messages that can seamlessly bridge different
standards or technologies. In other words, one of the main aims of
communication standardization for smart grids is ensuring
interoperability between different system components rather than
defining these components (meters, devices or protocols) \cite{20}.

In this context, generic application programming interfaces (APIs)
and middleware are useful enabling technologies. The success of
commercial deployment of smart metering and smart grid solutions
will significantly depend on the availability of open and standard
mechanisms that enable different stakeholders and vendors to
interoperate and interface in a standard manner. Open interfaces
serve many purposes and provide additional benefits in
multi-stakeholder scenarios such as smart energy management in home
and industrial environments. Further, open APIs provide the means
for third parties not directly associated with the original
equipment manufacturers to develop a software component which could
add functionality or enhancements to the system. On the other hand,
smart energy management solutions require access to more
information, ideally from different service providers and devices
implemented by different vendors. Such information should be
available and presented in a usable format to interested parties.
Timing and specific configuration of measurements and controls are
also critical for dynamic scenarios. Since support for different
technologies and some level of cooperation over administrative
boundaries are required, proprietary or widely simplified interfaces
will not be sufficient in these scenarios. This situation can be
improved by standard generic API definitions covering methods and
attributes related to capability, measurement and configurations.
The design of such APIs should be technology agnostic, lightweight
and future-proof.

\subsection{Scalable internetworking solutions}

Wireless sensor networks (WSN) research should be extended to smart
grid and metering. WSN has been an active research topic for nearly
ten years and has found many applications \cite{21}\cite{22}. Smart
grid/metering appears to be a major application for WSN, especially
related to Internet of Things and machine-to-machine (M2M)
communications \cite{23}. Existing industry efforts include IETF
6LoWPAN \cite{24} and ROLL \cite{25}\cite{26}\cite{89}. Based on
smart metering user scenarios, the overall M2M network architecture,
service requirements, and device capabilities are yet to be defined.
Recently ETSI has established a new M2M technical committee to
address these issues \cite{27}.

Internetworking between cellular networks and local area networks
(e.g. WLAN) has received a lot of attention because of the need for
seamless mobility and quality of service (QoS) requirements
\cite{28}. Topics such as intelligent handover and connection
management have been extensively investigated. In the context of
smart grid, due to the extremely large scale nature of the network,
the characteristics of the metering and control traffic carried in
the network are not clearly known. For instance, it could be the
case where 100,000 smart meters generate meter traffic data every 10
minutes. As a result, how to design and provision a scalable and
reliable network so that this data can be delivered to the central
utility control in a timely manner is a challenging task. As traffic
will be traversing different types of networks, interoperability is
the key. Further, some of the traditional research topics may need
to be revisited to cater for smart grid traffic, e.g. resource
allocation, routing, and QoS. This is because the traffic that will
be generated by e-energy type applications will likely be quite
different to the traditional browsing/downloading/streaming
applications that are in use today, with a mix of both real-time and
non-real-time traffic being generated and distributed across
different parts of a smart grid \cite{29}.

Interworking of communication protocols and dedicated smart metering
message exchange protocols such as DLMS/COSEM \cite{17} is an open
research issue. The DLMS/COSEM standard suite has been developed
based on two concepts: object modelling of application data and the
Open Systems Interconnection (OSI) model. This allows covering the
widest possible range of applications and communication media. Work
has already started in the industry trying to address the issue of
carrying DLMS data over various networks such as GPRS and power line
communications (PLC) networks (for an overview of PLC and its
applications to the smart grid, please refer to \cite{30}\cite{31}).
Recently, the DLMS User Association also established a partnership
with the ZigBee Alliance and the two organizations are working on
tunnelling DLMS/COSEM over ZigBee networks to support complex
metering applications. Inside IEEE 802.15.4, the 802.15 Smart
Utility Networks (SUN) Task Group 4g \cite{32} is working on a PHY
amendment to 802.15.4 to provide a global standard that supports
smart grid network applications with potentially millions of fixed
endpoints.

Recently a number of studies on efficient networking technologies
for energy management have been published in the literature. A
routing protocol for Advanced Metering Infrastructure (AMI) based on
the framework of IPv6 routing protocol for low power lossy networks
(RPL) is discussed in \cite{33}. An ETX (expected transmission
count) based rank computation method has been proposed for DAG
(directed acyclic graph) construction and maintenance in RPL. This
method enhances the unicast reliability of AMI networks. Simulation
results have shown that the proposed mechanism produces satisfactory
packet delivery ratio and end-to-end delay. In the near future,
electric vehicles (EV) are envisaged to be a major application in
the smart grid, and \cite{34}, among others, has outlined
architectures for vehicle-to-grid (V2G) communications. The
communication requirements and data flows between the EV and the
grid (with distributed energy source generators) have been
discussed, and Session Initiation Protocol (SIP) \cite{35} is
considered as a suitable solution for the establishment of
communications. V2G networks have also been studied in \cite{36},
focusing on the message structure (based on ISO/IEC 15118-2) and
message exchange sequence between the EV and the server over an
IPv6-based PLC communication network.

\subsection{Self-organizing overlay networks}

Because of the scale and deployment complexity of smart grids,
telecommunication network systems supporting smart grids are likely
to rely on the existing public networks such as cellular and fixed
wired access technologies, as well as private and dedicated networks
belonging to different administrative domains. The purpose of such
networks can be seen not only as a communications medium to exchange
monitoring and control information, but also as an enabler of new
services and applications. In many ways, the complexity and
heterogeneity characteristics of smart grid communications networks
will be similar to that of a wireless radio access network
supporting voice and data services. However, stakeholder
expectations, QoS requirements and load patterns will be
significantly different from those of a typical mobile voice/data
network because of the nature of the applications and services
supported. Both will share, at least partially, problems related to
managing and operating a complex and heterogeneous network where
tasks such as network planning, operation and management functions,
and network optimization are important. We believe that a
self-organizing network overlaid over existing infrastructure could
be the way forward to support wider deployment of smart grid
systems. Such a self-organizing network should support functions
such as communications resource discovery, negotiation and
collaboration between network nodes, connection establishment and
maintenance to provide the performance guarantees required by smart
grid/metering applications. Recently, novel network architectures
such as cloud-based systems have been proposed for smart grid data
collection and control \cite{87}\cite{88}.

\subsection{Home networking challenges}

Research on home networking has so far focused on providing
multimedia applications with high QoS, zero-configuration, and
seamless connectivity to home users. With the advent of smart grids,
new features and system design principles have to be considered. For
example, consider how to integrate smart meter or M2M gateway
functions into the home gateway (e.g. WLAN access point or femtocell
base station) in a cost-effective manner. Clearly, smart metering
adds a new dimension to home networks, complicating the issue of
interference management and resource provisioning.

With potentially every device and appliance in the home supplying
energy related information to the smart meter/home gateway (and by
extension, possibly to the energy supplier as well), it is easy to
envisage an order of magnitude increase in the number of devices in
each home that are able to communicate with each other and with the
outside world.  Today's homes may have 2-3 computers (desktop,
laptop, smart phone) that are connected to the home network and to
the Internet. Tomorrow's smart grid/smart meter homes could have
20-30 or more appliances and devices connected to the same network.
Although the preferred (wireless) networking standards for these
devices have not yet been established, it is clear that there will
be many more devices connected to whichever network is used.
Although there has been much discussion in the networking community
over the years of having ``an IP address for every possible device''
in the home, the convergence of energy provisioning and
communications may be the catalyst for this to actually become a
reality.

Along with the many new devices that will be connected to home
networks, new kinds of applications will undoubtedly emerge.  The
prime (and easy to envisage) application is the one of energy
consumption monitoring within the home and other areas (offices,
etc.).  In this direction, there are proposals for load monitoring
and real-time control from the utility companies' perspective.
However, energy monitoring has the potential to grow into something
far more significant than just measuring the energy consumed.  With
the current concerns over climate change and the very important need
for energy efficiency in all areas, it follows that fine-granularity
monitoring of energy usage in the home and other areas will become a
necessity and much research will be required in automating methods
for energy usage reduction in the home.  Given that there is much
perceived wastage in the way in which the appliances and devices are
used today (e.g. leaving devices on standby, inefficient usage of
washing machines and refrigerators, inefficient use of heating and
cooling), there is plenty of scope for realizing automated methods
for reducing energy consumption.

\section{Smart grid security and privacy issues}

Analyzing and implementing smart grid security is a challenging
task, especially when considering the scale of the potential damages
that could be caused by cyber attacks \cite{93}. For example,
protection against unauthorized access and repudiation is a vital
requirement for usage and control data communicated within the
system, assuming that critical system functionalities require that
the data are trusted by both the utility providers and the
customers. To provide such security services might not be trivial as
it may involve the integration of different information security
domains such as secure communication protocols, tamper-proof
hardware/software and regulatory frameworks on access control. The
need for protecting smart grid data cyber security emerges from the
need to interconnect smart grid components with a two-way
communication network so that energy suppliers and customers can
exchange information in an interactive, real-time manner. This
capability could assist in enabling features such as load shedding,
consumption management, distributed energy storage (e.g. in electric
cars), and distributed energy generation (e.g. from renewable
resources). Also, as previously discussed, the network could be
implemented using a variety of media ranging from fibre optic
broadband to ZigBee/WLAN, etc. Considering the need for fine-grained
monitoring of smart metering data, the security of an advanced
metering infrastructure is of paramount importance.

The security challenges of a smart grid system depend heavily on the
system architecture. For example, consider the smart metering
architecture in Figure \ref{sm}. Some logical components (such as
the distribution controller and the utility control services) may
communicate via a hard-wired tamper-proof link (e.g. implemented
within a simple physical controller) or may communicate using shared
networks (e.g. when there is physical fragmentation). Each case
imposes different security challenges such as protection against
single point failures or protection against multiple points of
attack. For example, the sink of a particular network can be
identified based on timing analysis of messages that are sent from
there \cite{74}. This could potentially allow an adversary to launch
an attack against the distribution controller which would have
serious impact on customers in the whole area. To this end, security
services should be provided through multiple layers of security so
that potential attacks result in minimal damages.

More generally, smart grid security risks and vulnerabilities can be
identified by using a top-down or a bottom-up approach. The top-down
approach analyzes well-defined user scenarios such as automated
meter reading (AMR) and billing, while the bottom-up approach
focuses on well-understood security attributes and features such as
integrity, authentication, authorization, key management, and
intrusion detection. A classification of smart grid risks and
vulnerabilities recently published by NIST can be found in \cite{37}
while a comprehensive specification of AMI security requirements has
been documented by OpenSG in \cite{38}.

\subsection{Cyber-physical security}

Smart grid cyber threats, such as the Stuxnet worm \cite{39}, have
the potential to breach national security, economic stability and
even physical security. Power stations and SCADA (supervisory
control and data acquisition) systems have always been targeted by
hackers. The move from closed control systems to open IP networks
opens up a new range of vulnerabilities. For example, data integrity
and authentication may be compromised through network attacks such
as man-in-the-middle spoofing, impersonation, or Denial of Service
(DoS) attacks. Similarly, data security may be compromised by
sabotage/insider attacks such as viruses and Trojan horses. The
latter threat becomes significant considering the potential openness
of the system and its interconnections with different networks such
as NANs and the Internet.

Once an entry point is found, it becomes easier for the attacker to
cascade an attack down the smart grid. For example, compromising the
real-time pricing channel may result in energy theft or malicious
remote control of appliances. Hence, rigorous hardware/software
security is required to ensure the validity of different
communicating parties such as head-ends and smart meters. If an
attacker takes over the head-end, then he might be able to send
smart meters a demand response command interrupting supply. The
interruption can be made permanent by also commanding all the meters
to change their crypto keys to some new value only known to the
attacker \cite{40}. The impact can be enormous: millions of homes
could be left without power until they are locally replaced or
rehashed with authentic keys, people suffer, health and safety could
be jeopardized, and businesses could lose millions. Smart grid
cyber-security needs to a) prevent such attacks from happening and
b) have a recovery/survivability mechanism in case of (successful)
attacks.

Communications security involves the design of a key management
crypto-system. This could for example be based on existing systems
such as Public Key Infrastructure (PKI) and Identity-Based
Encryption (IBE) \cite{41}\cite{42}. IBE, in particular, may be
attractive for smart grids as it can be deployed without prior
configuration, as the identity (ID) of a device is used to generate
unique keys. This allows easy deployment of low powered devices such
as sensors because they may start sending secure messages without
the need to contact a key server. In general, a mixture of
hierarchical, decentralized, delegated or hybrid security schemes
may be feasible. Preferably, a candidate scheme should include
secure bootstrapping protocols, i.e. it should provide effective
means to initialize new devices. Further, critical security
operations, such as key updates, should preferably employ group key
management techniques, such as {\it defence in depth} techniques
used in nuclear or military control systems, to mitigate the impact
of compromised head-ends (or trusted people).

For more information on different smart grid cyber security attacks
and threat impact, interested readers are referred to the NIST
guidelines \cite{37}.

\subsection{Privacy}

Frequent smart metering data collection and analysis can help
improve energy efficiency, as discussed in previous sections. Smart
meters are expected to provide accurate readings automatically at
requested time intervals to the utility company, the electricity
distribution network or the wider smart grid. However, this comes at
the cost of user privacy. That is, the information contained in such
data may be used for purposes beyond energy efficiency, which gives
rise to the smart grid privacy problem. In particular, frequent data
collection from smart meters reveals a wealth of information about
residential appliance usage, as discussed in \cite{44}.

In general, data privacy concerns the security of data that is
linked with, or infer information related to, the life of
individuals. The use of access control mechanisms, e.g. secure
authentication, authorization and confidentiality services, cannot
address the problem of smart grid data privacy holistically. This is
because these data need to be disseminated to many different
stakeholders within the grid. As discussed in \cite{75} the
consequences of the smart grid privacy problem are hard to
understand, because a) the full range of information extraction
possibilities are not known, and b) the concept of smart grid
privacy is still not well defined. A good reason of why the problem
of data privacy should not be underestimated can be found in a paper
on digital inclusion and its ramifications \cite{45}.

Currently, the smart grid privacy problem is highlighted by
Non-intrusive Appliance Load Monitoring (NALM) technologies that use
energy measurements to extract detailed information regarding
domestic appliance. Since the original work \cite{46} there has been
a wealth of research in the construction and upkeep of appliance
libraries and detection algorithms \cite{47}. Recent results suggest
that even when household power profiles are aggregated, the use of
household appliances can be identified with high accuracy \cite{48}.
The authors of \cite{49} have described a fine-grained energy
monitoring system that generates device level power consumption
reports primarily based on the acoustic signatures of household
appliances. Their experiments demonstrate that the system is able to
report the power consumption of individual household appliances
within a 10\% error margin.

\begin{figure*}
 \centering
   \includegraphics[height=4.6in]{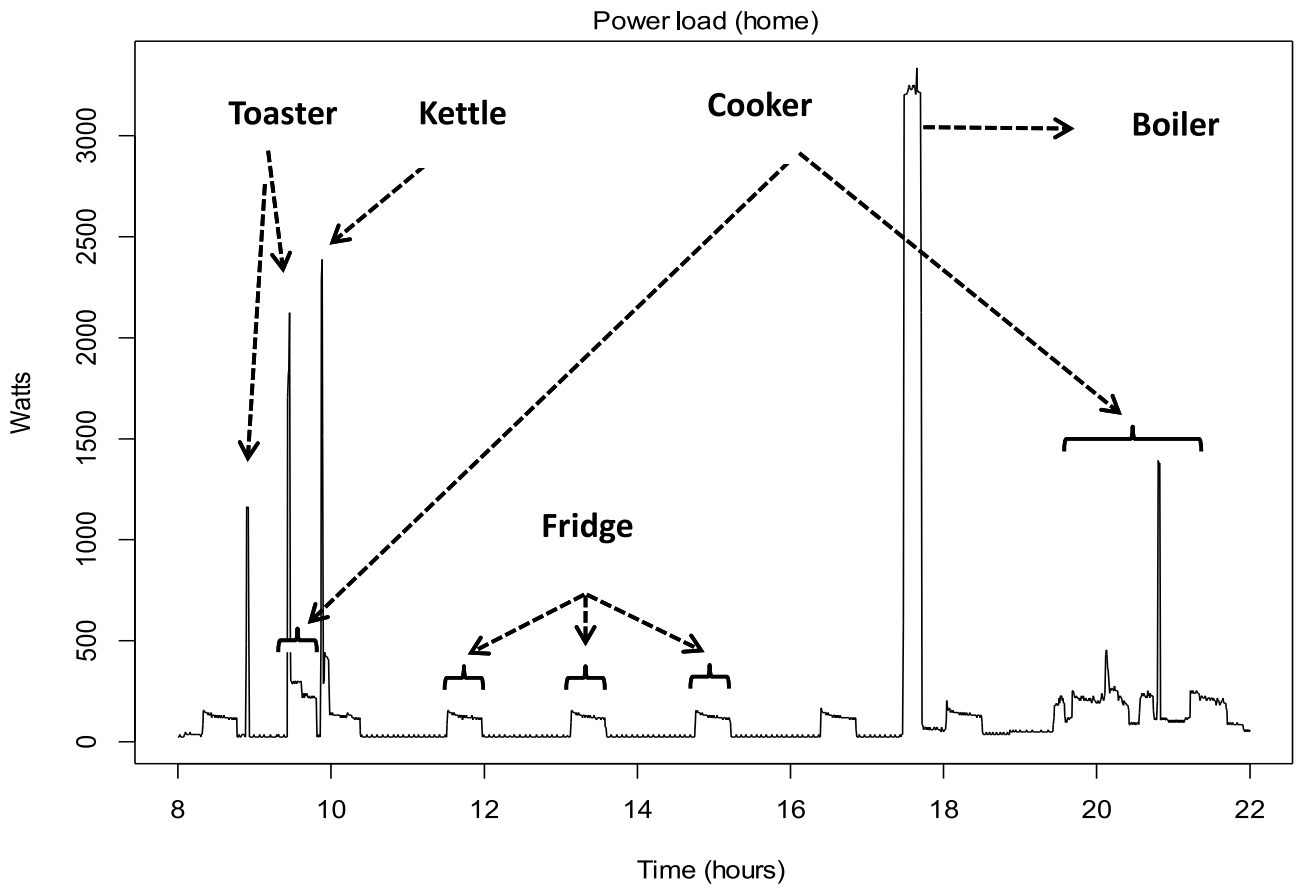}
\caption{Household electricity demand profile recorded on a
30-second time base from a one-bedroom apartment}
\label{load}
\end{figure*}

An example of appliance detection can be seen in Figure \ref{load}.
In general, the granularity of events that an algorithm may be able
to successfully detect depends on the frequency of smart meter
readings. The frequency range can vary, depending on the utility,
but in general this could be as high as every few (1-5) minutes.
Such detailed energy usage information could lay bare the daily
energy usage patterns of a household and enable deduction of what
kind of device or appliance was in use at any given time. Further,
the authors in \cite{76} suggest that data mining techniques can be
used to reveal trends of personal behavior in the metering data even
if relatively low data sampling rates (e.g. every 30 minutes) are
assumed.

From the above it becomes clear that the smart grid privacy problem
is important and solutions are needed. There are two classes of
privacy protection schemes: a) regulatory-based ones and b)
technological-based ones. Current standardization activities focus
on developing regulations and policies to help protect smart grid
privacy. In the USA, NIST has acknowledged that the major benefit
provided by the smart grid, i.e. the ability to get richer data to
and from customer meters and other electric devices, is also its
Achilles' heel from a privacy viewpoint \cite{37}. Further, the
American National Association of Regulatory Utility Commissioners
(NARUC) \cite{43} has drafted a resolution stating that: ``utility
customer information can be used to differentiate utility services
in a manner that creates added value to the customer''. On the other
hand, ``a balance has to be carefully considered between the
appropriate pro-competitive role that utility customer information
can play in new and developing markets and the privacy implications
of using that information''.

In Europe, the European Commission has set up a Task Force on smart
grid aiming to develop a common EU smart grid vision and identify
key issues that need to be resolved. In response, three Expert
Groups (EGs) have been set up, one of which, EG2, aims to identify
the appropriate regulatory scenarios and recommendations for data
handling, security and consumer protection. One of the EG2
recommendations is to use anonymity services to help protect
privacy. For example, smart metering data can be separated into low
frequency attributable data (e.g. data used for billing) and high
frequency anonymous technical data (e.g. data used for demand side
management). In this case, the main challenge resides in anonymizing
the high-frequency data, which are required for efficient grid
functionalities, while making sure that the reliability, the
effectiveness and the security of these functionalities are not
compromised.

A number of technological solutions have further been proposed as
follows.
\begin{itemize}
\item Anonymization.  An example of this direction is studied in
\cite{50} where a secure escrow protocol is proposed to securely
anonymize the ID of frequent metering data sent by a smart meter.

\item Aggregation. In \cite{51} the authors introduce two different solutions
for the smart grid data privacy model. One solution uses a third
trusted party as a data aggregation proxy. The other solution adds
random value from a particular probability distribution to the data.

\item Homomorphism. The use of homomorphic encryption can corroborate
smart meter data privacy as discussed in \cite{77}.  This paper
develops a method for a number of meters that have a trusted
component and enjoy a certain level of autonomy. A trustworthy
system provides guarantees about the measurements for both grid
operators and consumers.

\item Obfuscation. In \cite{78} a cooperative state estimation
technique is introduced that protects the privacy of users' daily
activities. The proposed scheme can obfuscate the privacy-prone data
without compromising the performance of state estimation.

\item Negotiation. In \cite{79} the authors introduce the concept of
competitive privacy between the utility that needs to share the data
to ensure network reliability and the user that withholds data for
profitability and privacy reasons. The resulting trade-off is
captured using a lossy source coding problem formulation.

\item Energy management. In \cite{53} the authors introduce a battery
management algorithm that changes home energy consumption in a
manner that helps protect smart metering privacy.

\end{itemize}

Although there is still much more research to be done in this area,
it appears that smart grid privacy is a sensitive topic which can be
approached in a number of different angles. Future protection
schemes is likely to be a combination or evolution of the solutions
introduced above, depending on the system cost and the need for
privacy in different societies.

\subsection{Secure integration}

The most widely discussed smart grid security challenges concern the
protection of smart metering data against unauthorized access and
repudiation. This is an important requirement without which AMR data
will not be trusted by either the utility providers or the
customers. Solutions are required on different levels: end to end
secure communication protocols need to be used, hardware components
(e.g. smart meters) need to withstand physical attacks, the grid
needs to detect forged/hacked components, and smart meter software
should be bug-free \cite{52}.

We believe that AMI communications security requirements can be
addressed by combining existing cryptographic protocols and
tamper-proof hardware solutions, by exhaustively testing equipment
and software against all sorts of attacks, and by adopting an open
architecture for further testing and secure updating. Also, as
discussed in the previous section, it is equally important to
develop mechanisms for protecting smart metering data against
insider attacks. The use of open smart grid interfaces, as
previously described, will create a gateway for multiple third
parties (stakeholders) to access and process AMR data. We need to
make sure that such insiders will access smart metering data in an
authorized manner and will only use this data in an {\it acceptable}
manner.

Our vision is that security policies and legislation are not a
panacea for privacy as they do not thwart attacks such as data
privacy concessions: history teaches that legitimate techniques for
mining and exploiting data evolve quickly when there is a clear
financial incentive. Hence, the problem of smart meter data access
and usage needs to be further reviewed within different security
domains:

\begin{itemize}
\item Enforcement: Smart metering data should belong, in principle, to the users. For example, a digital
rights management system could be used to allow utility providers to
use the data in an acceptable manner. Any use of personal data
(acceptable or unacceptable) should not be repudiated.

\item Concealment:  Users may use power routing algorithms in a
manner that conceals events extracted from their energy usage
activities \cite{53}.

\item Reaction: There should be mechanisms that will detect (in
retrospect) misuse of smart metering data. These mechanisms should
have regulatory support for counter-measures (e.g. penalties)
against malicious parties.

\end{itemize}

The common challenge in all the above cases remains to design a
system that will balance the trade-off between security and
performance, i.e. use adequate security strength while minimizing
its power usage and cost overheads. In the future, smart
grid/metering communications may potentially integrate with
heterogeneous network systems, Internet applications, etc. For
example, a roaming smart grid customer may wish to use power energy
in remote areas and link this usage to his smart metering profile.
This integration can be used for billing or other personalized
services. In such an example, it will be necessary to establish
secure communication protocols between different parties such as a
home smart meter, a mobile phone, a smart roaming power appliance
and the customer. The customer may additionally wish to allow third
parties get access to personal smart metering information in
exchange for services such as free entrance to facilities, or the
customer may wish to remain anonymous. One can envisage further
challenges arising as smart grid communication systems integrate
with other communication systems: home entertainment systems,
medical communication systems, and traffic monitoring communication
systems (e.g. via GPS), just to name a few.

In the above scenarios it becomes clear that integration of services
and interfaces gives rise to a whole new range of security and
privacy vulnerabilities and requirements. In such complex computing,
communications and energy management environment, it is important to
understand how security risks are cascaded, i.e. how the compromise
of one system leads to compromise of a downstream system. Risk
analysis should be able to detect both proactive and reactive system
anomalies and take appropriate measures such as creating appropriate
logs and alerts. Further, the extrapolation and combination of
multi-domain information such as energy consumption data, location
information, lifestyle information, and other personal information
increases the potential both for richer applications and services as
well as security threats and damages. Indeed, future integration of
systems and services requires transparent and secure protection
mechanisms, more than ever before.

\section{Energy management - building on lessons learnt from communications
networking research}

In previous sections we have discussed research challenges for
communication engineers posed by the smart grid. On the other hand,
technologies developed in communication networking can help to solve
various problems in energy management too. We elaborate on this
interesting topic in this section. One of the main problems
currently faced by the energy suppliers is the fluctuations in
energy demand, which are expected to be further exacerbated when
plug-in electric vehicles become a reality in the near future. In a
situation like this, the approach traditionally adopted by the
energy supplier is to take the peak consumption into account and
create enough reserve energy supply (potentially upgrade existing
infrastructure) to meet this uneven energy demand. This is similar
to the over-provisioning approach adopted by the communications
network service providers. Even though this approach addresses the
problem of meeting the demand, it results in inefficiencies by
creating a waste in the system since the demand on average is much
less than the (estimated) peak energy requirement. Installing power
plants to cater for the peak demand is not only expensive, but also
may not be practical as the demand often keeps outpacing the supply.
A coordinated effort to shift the power load from peak to off-peak
times will lead to better generator utilization and fewer standby
sources of energy which translates into reduced cost to the utility
provider and less damage to the environment.

A compilation of energy consumption statistics of typical household
electrical appliances can be found in \cite{54}. This data suggests
that most of the hard hitters, e.g. storage heaters, dish washers,
washing machines, etc. can be rescheduled to operate during the odd
hours, e.g. after midnight or early morning, when the overall demand
is low. The next step is to identify how to determine the demand at
the consumer level, distribution system level, and the aggregate
demand taking into account all the distribution systems. This
requires bi-directional communication mechanisms between the grid
and the consumer premises - facilitating the consumers to notify
their demand requirements to the grid and for the grid to feedback
availability/pricing information to the consumers. Based on the
information obtained from the consumer side, the utility providers
can ascertain demand per distribution system and hence the aggregate
demand. Subsequently, considering the capacity and the demand, the
utility provider may identify target operating points for each
distribution system. These could then be conveyed to the
distribution systems following which the distribution systems could
then translate these into individual targets for the consumers
connected to it. Such target operating points clubbed with the
consumer preferences could then factor into the decision of load
scheduling at the consumer premises.

We can look at this problem as one involving balancing the load so
as not to exceed the available capacity which, in essence, is
similar to the traffic engineering problem in the Internet where
traffic load balancing is analogous to electrical load balancing. We
observe parallels between the two problems in that, mechanisms to
estimate the demand are analogous to available bandwidth estimation
and the concept of a traffic engineering management server spreading
traffic across different paths is analogous to the energy management
system distributing the available supply appropriately to the
different distribution systems. Common to both are concerns of
fairness (i.e. how to share the available capacity amongst the
different users/consumers?) which could build upon the numerous
studies on fairness in the communications networking area.

\begin{figure*}
 \centering
   \includegraphics[height=3.0in]{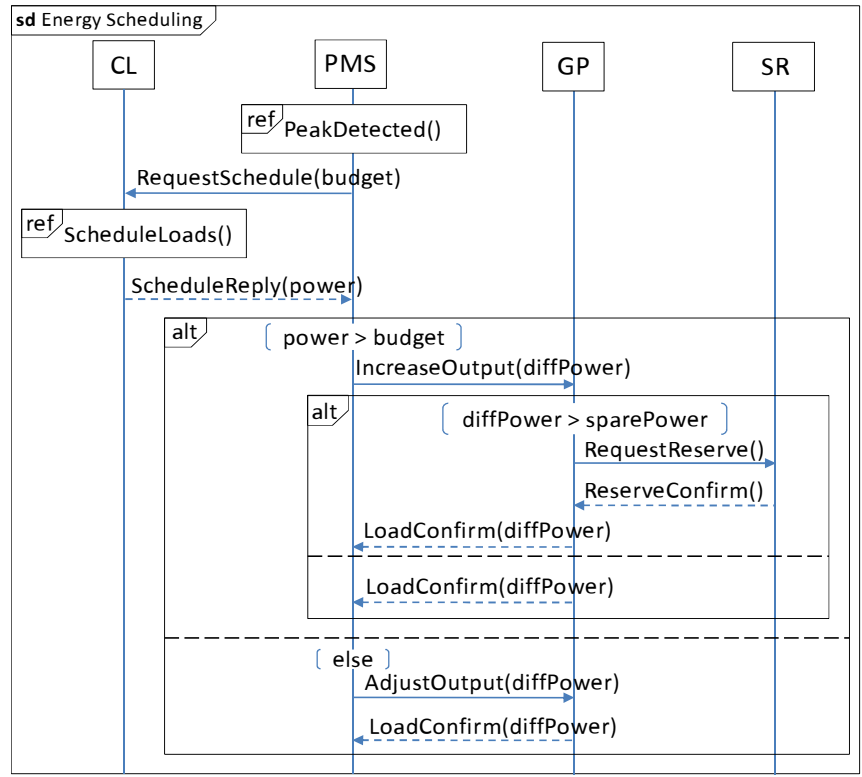}
     \vspace{30mm}
\caption{Example of an energy management scenario} \label{sedat}
\end{figure*}

On a similar note, we observe that the problem of meeting the
operating targets assigned to consumers is similar to the resource
scheduling problem; essentially how to schedule the different
devices within the house so that the net consumption conforms to the
operating target. Figure \ref{sedat} \cite{55} depicts the
interaction between a power management system (PMS), aggregate
consumer load (CL), generation plant (GP), and spinning reserve (SR)
when a peak is detected in a smart grid. The aggregate power
consumed by the PMS customers needs to be brought into a power
budget (budget) assigned to the PMS by the smart grid. When a peak
event is detected (e.g. certain threshold being breached), the PMS
could send its customers a schedule request to match the demand to
the available supply by rescheduling non-critical loads to off-peak
periods. This could possibly result in some reduction in demand. If,
however, the demand still exceeds the allocated power budget, the
PMS could send a request to the GP to increase its output by the
amount of difference between the re-scheduled power demand and the
budget of the PMS (diffPower). It may be likely that the GP may not
be able to fulfil this demand in which case it could request the SR
to allocate extra power to the GP to meet the power demand of the
PMS customers. On the other hand, if the demand (subsequent to
customers rescheduling some of their loads) drops below the budget
allocated to the PMS, a request to reduce the power output could be
sent to the GP in order to conserve energy. This ensures better
utilization of the available resources by simply shifting power
demand to the low demand periods thereby leveling the load.

Finally, it may be likely that it is impossible to satisfy the
demand in which case the energy supplier may have to resort to
controlled (partial) outage. The decision of who does and doesn't
encounter outage could be based on the different preferences and
priorities of the different consumers depending on their roles in
the grid. It is envisaged that the prioritization could be done in a
hierarchical manner wherein the consumers are grouped into different
service classes at the time of registration (initiation of service).
For example, hospitals/emergency services should never encounter
outage whenever possible, whereas service to other low priority
consumers may be compromised. Load shedding should happen in such a
way so that loads are disconnected according to the service classes
assigned to them. The consumer in each service class could further
prioritize their appliances according to their preferences which
help the scheduler to make scheduling decisions for the consumer in
an outage event where the outage can be avoided by merely
re-scheduling non-critical loads in the lowest priority service
class (e.g. residential). This is analogous to the QoS paradigm in
computer networks where some of the traffic is dropped or delayed
(using the priority assigned to the packets) due to lack of
bandwidth. Different classes of priorities could be assigned to
different types of buildings according to their perceived
importance. Further, having sub-classes can facilitate more granular
load scheduling decisions.

Apart from the aforementioned approaches, we also observe a number
of similarities between the smart grid resource allocation problem
and well-known problems from communications and networking research.

{\it Load Admission Control}: One of the popular mechanisms for
resource management in the communications networking world is
admission control wherein a flow/connection is admitted in the
network only if resources are available to serve it. Intuitively,
this could be an interesting approach to explore in the context of
energy management. This could potentially work as follows: the
utility could allocate target operating points to each distribution
system connected to it. Prior to connecting the load, the energy
management gateway at the consumer's premises could send a load
request to the Power Management System connected to its distribution
system. This could contain both the load that is desired and the
priority of the load. If the sum of the incoming load request and
the current aggregate load is less than the target operating point,
the load request could be granted. Otherwise, if the priority of the
load request is low, a response could be sent to the consumer to
reschedule the load. If the priority of the load request is high,
then the PMS could send a message to the noncritical loads connected
to it, asking them to respond if any of them would be willing to be
rescheduled. Once the responses are received, it could choose the
first of the responses which can relinquish load equivalent to the
demand of the pending request. The PMS could then instruct this
consumer to reschedule (providing a cheaper tariff as a reward) and
approve the load request which is pending. After approving the
pending request, the PMS would update its current aggregate load by
adding the recent request's requirement.

As elaborated in the earlier sections, the priority of different
loads will vary depending on the individual customer preferences.
Hence, it should be possible to adjust the priority of each load
according to the customer's perceived priorities. This can be
achieved using the home energy management gateway (HEMG) device
which will provide the necessary functionality to assign preferred
priorities to customer loads. Although load admission control
requires more information exchange between HEMG and PMS which may
raise privacy concerns, this exchange can be anonymized using
privacy measures such as that introduced in \cite{50}.

{\it Load Scheduling}: There are similarities between the load
scheduling problem in smart grids and scheduling and resource
allocation problems in communication networks. For example, in
Orthogonal Frequency Division Multiple Access (OFDMA) wireless
systems \cite{56}, mathematical optimization algorithms are used to
determine optimum power allocation and frequency band selection to
satisfy different data rate requirements for various users. This has
similarities in smart grids where loads need to be allocated to
various users at different time slots to optimize resource
utilization and to improve energy efficiency. The solutions to load
balancing in smart grids can be obtained based on the optimization
approaches used in communication networks.

{\it Cooperative Energy Trading}: It is envisaged that a future
smart grid will include many micro generation plants. We envisage a
cooperative energy trading scheme where the energy trading happens
between energy users in a local open market. A local market could
consist of microgrids that operate their own micro generation plants
as well as being connected to the macrogrid \cite{57}. The idea here
is a better utilization of the available power resources by
cooperatively using available generation resources. This approach is
very similar to the cooperative communication \cite{58} philosophy
where the nodes in a wireless network try to increase their
throughput and network coverage by sharing available bandwidth and
power resources cooperatively. The consumers with micro generation
plants create a market for trading energy within their local
consumer group. They cooperatively use energy generated by the local
micro generation plants to curb their dependence on the macrogrid
supply. This approach will both increase the efficiency of the
macrogrid and possibly fetch a better price for micro generation
electricity. If the micro generation output in a cooperative energy
area is greater than the total power consumption (surplus power),
the same cooperative trading concept can be applied to the
microgrids that are in close proximity (e.g. neighboring
microgrids). This approach could greatly reduce the transmission
costs by localizing energy distribution in a hierarchical manner.

To summarize, the lessons learned and the optimization approaches
adopted in communication networks can be used in energy management
problems in smart grids. While solutions to resource allocation in
communication networks are well established, adopting these within
the context of smart grids may not be trivial. Nonetheless, this
could be a good starting point as these concepts could potentially
offer invaluable insights into the design of reliable and efficient
smart grids \cite{55}. There has been a recent surge of interest in
extending the optimization approaches used in communication networks
to demand management in smart grids. For example, the work in
\cite{59} uses convex optimization techniques for the power
consumption scheduling of appliances to minimize peak-to-average
load ratio. The concept of congestion pricing in Internet traffic
control has been applied to smart grid demand response in \cite{60}.
The work in \cite{60} has demonstrated that the burden of load
leveling can be shifted from the grid (or supplier) to the end users
via this pricing mechanism. This idea has been further extended to
distributed electric vehicle charging in \cite{61}. A least-cost
dispatch of available generation resources to meet the electrical
load has been proposed using a unit commitment mode that relates the
demand side and supply side management through a hidden Markov model
and a Markov-modulated Poisson process in \cite{62}. Most recent
work in this area can also be found in, e.g.,
\cite{83}\cite{84}\cite{86}.

We finish this section by briefly discussing some of the challenges
of demand side management (DSM). Having surveyed various
optimization techniques for DSM, it should be highlighted that
regardless of the potential economic benefits, not all consumers
will be willing to participate in demand side management due to
reasons for example the consumption discomfort introduced by load
scheduling. Therefore, more work in terms of incentive design is
required for the successful deployment of coordinated scheduling
algorithms. Further, various distributed energy generation and sell
back technologies can help further reduce the dependency on central
supply. Therefore, optimization of energy dispatch at residential
level is also a significant challenge for future DSM.

\section{Smart metering standardization activities in Europe}

To realize the vision of smart grid enabled by technologies
mentioned above, it is essential that we have a whole set of
industrial standards in place to ensure interoperability and
reliability. A key component of smart grid is the smart meter, which
is capable of performing detailed measurements at customer premises
and reporting them back to the utilities. Smart meters and the
information they generate will provide the glue that allows the
components of a smart grid to work together effectively and
efficiently. A major paradigm shift in the operation, management and
behavior of the energy industry could be achieved through smart
metering.  With the target of near-real-time response to potential
grid problems, the communication flow between millions of
residential customer premises will also be bi-directional and at
many levels.  Examples include user load controls, smart meters,
user distributed energy source control (i.e., micro-generation at
the home), home network and energy control, power distribution
equipment, supply and demand control software, large-scale power
source control (i.e. power stations), and of course, billing and
service provisioning infrastructure at the utilities themselves. For
instance, having access to real-time information on the flow of
energy in the grid enables utilities and consumers to make smarter
and more responsible choices.  Also, the ability to monitor demand
and supply allows remote sensing of damages, fault detection and
tampering (electricity theft).  At the same time, significant
opportunities are presented to technology solution providers to
develop enablers and techniques to deal with complex supply, demand
and controls and to facilitate sustainable energy production and
security.

While the benefits of smart meters will be common across the world,
their functionality, the adopted communications technologies,
standardization and regulation will be different due to
geographical, economical, political and social factors.    The
envisaged systems are highly complex, not only due to the high level
of distribution but also due to the large number of stakeholders
with direct interests in the process.  Therefore, a key enabling
role to ensure successful integration is to be played by standards,
as clearly recognized in \cite{63} and \cite{19}.

Following the third Energy Package, Directive 2009/72/EC of 13 July
2009, the European Commission has decided to set up a Task Force on
smart grids aiming to develop a common EU smart grid vision and
identify key issues that need to be
resolved\footnote{http://ec.europa.eu/energy/gas\textunderscore
electricity/smartgrids/taskforce\textunderscore en.htm}. The Task
Force consists of a steering committee and three Expert Groups. The
high level steering committee includes regulatory bodies,
Transmission Systems Operators (TSOs), Distribution System Operators
(DSOs), Distribution Network Operators (DNOs), and consumer and
technology suppliers working jointly to facilitate the smart grid
and smart metering development, and supporting the achievement of
the 2020 targets. The three EGs are the following:

\begin{itemize}
\item EG1. Functionalities of smart grids and smart meters (current
state of the art, services, smart grid components, functions,
strategy for standards);

\item EG2. Regulatory recommendations for data safety, data
handling and data protection (need for standardized data model,
cyber security);

\item EG3. Roles and responsibilities of actors
involved in the deployment of smart grids, such as DSOs/DNOs and the
role of standards.

\end{itemize}

It is clear that there is a conscious effort within Europe to
harmonize smart metering/grid standards, and to create a single set
of European standards that will be widely used.  A significant part
of this endeavor targets communications architectures and solutions.
This section summarizes the European perspective on smart metering.

\subsection{Smart metering communication standardization in Europe}

Metering standardization, including automatic/remote meter reading,
is a well-established activity both at international and European
standardization levels (CEN/CENELEC and ETSI).  However, the
functionality envisaged for smart meters (e.g. support for multiple
dynamic tariffs/time of use tariffs, energy export functions,
variable scheduled meter reading, demand control, etc.) requires
interfacing to and either adopting existing or establishing new data
formats and standards in areas not considered by the metering
community so far.  This has required bringing together and
streamlining activities from multiple technical committees and
standardization organizations for smart metering.

In response to the identified need for comprehensive standards for
all aspects of smart grids, the European Commission issued Mandate
M/441 EN \cite{19} in March 2009.  This mandate was targeted to the
standardization bodies of CEN (European Committee for
Standardization), CENELEC (European Committee for Electrotechnical
Standardization) and ETSI (European Telecommunications Standards
Institute), with the prime objective being ``...to create European
standards that will enable interoperability of utility meters
(water, gas, electricity, heat), which can then improve the means by
which customers' awareness of actual consumption can be raised in
order to allow timely adaptation to their demands (commonly referred
to as smart metering)''. The current target for the European
standard for smart meter communication solution is August 2010, with
the harmonized solution for additional functionalities being
completed by December 2011.

Table \ref{standard1} and Table \ref{standard2}\footnote{Please note
that the connections we show in Figure \ref{sm} are physical
connections, whereas Table \ref{standard1}, Table \ref{standard2}
and Figure \ref{sm2} refer to standardization efforts and are more
to do with the logical connections between entities in the smart
metering infrastructure.} present the various communication and data
exchange standards applicable in different segments of the
end-to-end smart metering distributed system. More specifically, we
have grouped together standards, either agreed or proposed in the
relevant technical committees (TCs) in European standardization
organizations (CEN/CENELEC and ETSI), covering the HAN and WAN
elements of the communication system supporting the smart metering
applications, and layered according to the application, network and
transport, and communication link technologies.


\begin{table}
  \begin{center}
\begin{tabular}{ | p{1.2cm} | p{1.4cm} | p{1.4cm} | p{1.4cm} | p{1.4cm} |} \hline\hline


 & Connection between smart meters, devices, and displays & SM to SM-GW & HBES (home automation
network device, control, server, external server) & HBES to SM or
SM-GW interface \\
\hline

Application & ZigBee Smart Energy 1.0/2.0, proprietary data model &
{\bf CEN TC 294}: EN 13757-3 M-Bus, {\bf CLC TC 13}: IEC 62056 COSEM
& {\bf CLC TC 205}: EN 50090-3 & {\bf CLC TC 205}: EN 50090-3, {\bf  CLC TC 13}: IEC 62056 COSEM \\
\hline

Network and transport & ZigBee 2.0, 6LoWPAN  & ZigBee 2.0, 6LoWPAN &
{\bf CLC TC 205}: EN 50090-4 & {\bf CLC TC 205}: EN 50090-4 \\
\hline

Link and physical media & ZigBee, PLC, 802.15.4, Bluetooth,
Proprietary  protocols & {\bf CEN TC 294}: EN 13757-2 M-Bus wired,
{\bf CEN TC 294}: EN 13757-4 M-Bus wireless, {\bf CLC TC 13}: IEC
62056-31, Euridis 2 & {\bf CLC TC 205}: EN
50090-4 & {\bf CLC TC 205}: EN 50090-4 \\
\hline

\end{tabular}
  \end{center}
  \caption{HAN communication and data exchange standards for smart
metering in Europe}
    \label{standard1}
\end{table}

\begin{table}
  \begin{center}
\begin{tabular}{ | p{1.2cm} | p{1.6cm} | p{1.4cm} | p{1.6cm} |} \hline\hline


 & SM-GW to Data Concentrator & Concentrator to DCS  & SM-GW to DCS \\
 \hline

Application & {\bf CLC TC 13}: IEC 62056 COSEM & SMTP, SFTP, Web
Service, COSEM & {\bf CLC TC 13}: IEC 62056 COSEM \\
\hline

Network and transport & TCP/IP  & TCP/IP &
TCP/IP \\
\hline

Link and physical media & {\bf CLC TC 13 / IEC TC 57}: IEC 62056
COSEM, DLMS/COSEM over IEC 61334/S-FSK, PLC, GPRS and/or
Ethernet/ADSL & GPRS/GSM, PLC G3, Fibre VLAN, Point to multi-point
radio & {\bf CLC TC 13 / IEC TC 57}: IEC 62056 COSEM, DLMS/COSEM
over GPRS
 \\
\hline

\end{tabular}
  \end{center}
  \caption{WAN communication and data exchange standards for smart
metering in Europe}
    \label{standard2}
\end{table}

Currently the HAN area is populated strongly by IEEE 802.15.4-based
communication technologies, and in particular ZigBee-based
solutions, with the forthcoming ZigBee v2.0 with native support for
6LoWPAN \cite{24} aiming to provide seamless IP networking
connectivity between the smart meters, the metering gateway and the
home appliances. The other candidate technologies are Bluetooth, but
with no specific provisioning for application support for energy
management (which ZigBee currently provides through the ZigBee Smart
Energy Profile \cite{64}), and various narrow-band power-line
communication solutions.  However, it is to be noted that the
currently existing standards for full-protocol stack communication
in a HAN for home automation applications (Home and Building
Electronic Systems - HBES) are already proposed as the way forward,
and are currently covered by the CEN/CENELEC TC 205 technical
committee in the EN 50090-x series.

The WAN solution is largely populated in the data exchange format
(application) layer through the standards output from CEN/CENELEC TC
13 and the IEC TC 57 committees. Analyzing the proposed solutions
and the existing standards, it is obvious that an IP-based solution
at the network layer will have lower integration and
interoperability costs than any non-IP based solution.

While Table \ref{standard1} and Table \ref{standard2} mainly focus
on communication network standards for smart metering, there are
other important standards that cover different aspects of smart
metering. For example, IEC 61968-9 specifies interfaces for meter
reading and control. It specifies the information content of a set
of message types that can be used to support many of the business
functions related to meter reading and control, e.g. meter reading,
meter control, meter events, customer data synchronization and
customer switching. It also defines a list of functionalities such
as metrology, load control, demand response and relays, as well as a
related set of XML-based control messages \cite{65}.

\subsection{Standardization technical committees}

\subsubsection{Smart Meters Coordination Group (SM-CG)}

The SM-CG was set up as a Joint Advisory Group between CEN, CENELEC
and ETSI to manage the standardization work in support of European
Commission Mandate M/441 for the creation of European standards for
an open architecture for utility meters enabling interoperability
and to improve customer awareness of consumption.  The SM-CG is not
empowered to develop standards, but to propose allocation of the
work to existing CEN, CENELEC and ETSI technical committees.

\begin{figure*}
\vspace{25mm}
 \centering
   \includegraphics[height=7.0in]{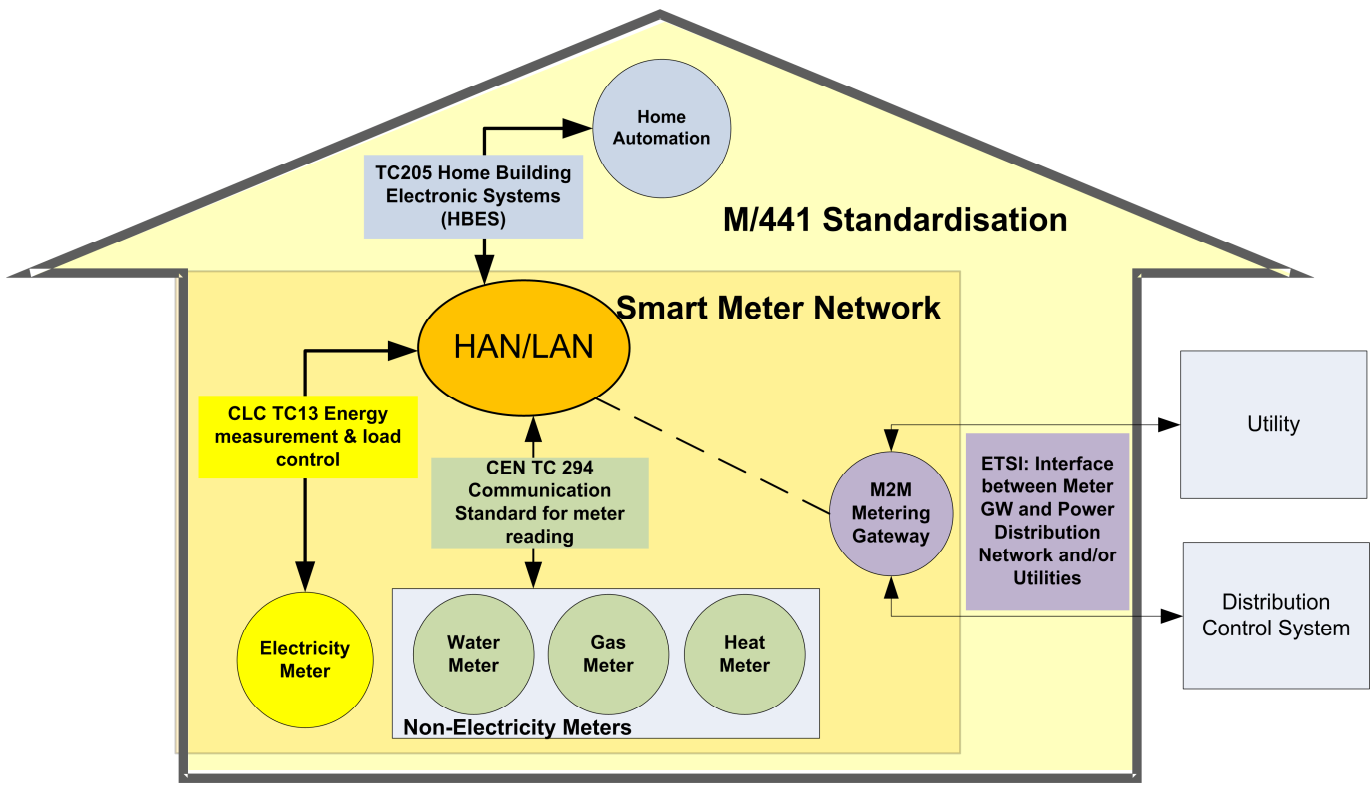}
   \vspace{-100mm}
\caption{TC responsibilities in European smart metering
standardization} \label{sm2}
\end{figure*}

As can be seen in Figure \ref{sm2}, the European perspective on
smart metering (especially within the home area) has been split into
three distinct standardization targets: electricity meters,
non-electricity meters (gas, water and heat) and home automation
\cite{20}. The M/441 standardization area also includes an M2M
remote gateway which will send the collected metering data to the
wider network (to be used by utilities and other interested
parties). Beyond basic smart metering and communications
functionality, a number of additional functionalities (F1 to F6)
have been identified as optional (but strongly desirable) features
of smart meters:
\begin{itemize}
\item F1 - Remote reading of metrological register(s) and provision of
these values to designated market organizations,

\item F2 - Two-way communication between the metering system and
designated market organizations,

\item F3 - Meter supporting advanced tariffing and payment
systems,

\item F4 - Meter allowing remote disablement and enablement of
supply,

\item F5 - Communicating with (and where appropriate directly
controlling) individual devices within the home/building,

\item F6 - Meter providing information via portal/gateway to an
in-home/building display or to auxiliary equipment.

\end{itemize}

\subsubsection{CENELEC TC 13 - Equipment for electrical energy measurement and load
control}

The scope of CENELEC TC 13 is to prepare European Standards (using
whenever possible IEC standards) for electrical energy measuring and
electrical load control equipment (such as watt-hour meters,
var-hour meters, maximum demand indicators, telemetering for
consumption and demand, equipment for remote meter reading, time
switches, equipment for the control of loads and tariffs and
consumer services) including the equivalent electronic forms of
these devices and their accessories. The activities of TC 13
encompass the development of the IEC 62056 DLMS/COSEM.

\subsubsection{CEN TC 294 - Communications systems for meters
and remote reading of meters}

The work of CEN TC 294 encompasses the standardization of
communication systems for meters and remote reading of meters for
all kind of fluids and energies distributed by network and not
necessarily limited to household meters. The activities of TC 294
encompass the development of the EN 13757 suite of standards on
Meter-Bus (M-Bus) and wireless M-Bus.

\subsubsection{CENELEC TC 205 - Home and Building Electronic
Systems}

The scope of CENELEC TC 205 is to prepare standards for all aspects
of home and building electronic systems in relation to the
Information Society. In more detail: to prepare standards to ensure
integration of a wide spectrum of control applications and the
control and management aspects of other applications in and around
homes and buildings, including the gateways to different
transmission media and public networks taking into account all
matters of EMC (electromagnetic compatibility) and electrical and
functional safety. TC 205 will not prepare device standards but the
necessary performance requirements and necessary hardware and
software interfaces. The standards should specify conformity tests.
TC 205 will perform the work in close cooperation with relevant
CENELEC TCs and those in CEN and ETSI. The activities of TC 205
encompass the development of the EN 50090 and EN 50491 suites of
standards on Home and Building Electronic Systems (HBES) and
Building Automation and Control Systems (BACS).

\subsubsection{ETSI M2M}

As mentioned previously, a new ETSI Technical Committee is
developing standards for machine to machine communications.  The
group aims to provide an end-to-end view of machine to machine
standardization, and will cooperate closely with ETSI's activities
on Next Generation Networks, and also with the work of the 3GPP
standards initiative for mobile communication technologies.
According to the Terms of Reference for ETSI TC M2M \cite{66}, the
responsibilities of the ETSI TC M2M are the following:

\begin{itemize}

\item ``to collect and specify M2M requirements from relevant stakeholders;

\item to develop and maintain an end-to-end overall high level
architecture for M2M;

\item to identify gaps where existing standards do
not fulfil the requirements and provide specifications and standards
to fill these gaps, without duplication of work in other ETSI
committees and partnership projects;

\item to provide the ETSI main center of expertise in the area of M2M;

\item to coordinate ETSI's M2M activity with that of other
standardization groups and fora.''

\end{itemize}

As part of their activities in smart metering, ETSI M2M has recently
approved a smart metering use cases document \cite{27} which
discusses various detailed use cases in relation to a typical smart
metering configuration, as shown in Figure \ref{sm}.  Examples of
these use cases include: obtain meter reading data, install,
configure and maintain the smart metering information system,
support prepayment functionality, monitor power quality data, manage
outage data, etc. All these use cases are discussed in terms of
general description, stakeholders, scenario, information exchanges,
and potential new requirements.

The use case of ``obtain meter reading data'', as an example, is
aligned to the additional functionalities F1 and F2 as described
previously. It describes how the Smart Metering Information System
provides this meter reading data to the Read Data Recipient, either
periodically or on request. The stakeholder for this use case is
Read Data Recipient. The information exchanges involve both Basic
Flow and Alternative Flow. For instance, the basic flow for
automatically scheduled readings is: Smart Metering Information
System registers meter reading along with time/date stamp, and Smart
Metering Information System sends meter reading data to Read Data
Recipient. The potential new requirements (from an M2M perspective)
for this use case are mainly regarding reliability and security:
minimal latency for on-demand readings of distribution network
management applications such as overload or outage detection;
accurate and secure time synchronization; support of various
information exchanges from M2M Devices and M2M Gateways;
authentication of the M2M Device or M2M Gateway; a security solution
in place to prevent eavesdropping at any point in the network;
verification of the integrity of the data exchanged \cite{27}.

A number of liaisons have also been established with other
standardization bodies, e.g. CEN, CENELEC, DLMS UA, ZigBee Alliance
and other ETSI TCs. The TC M2M domain of coordination to answer
M/441 includes: providing access to the meter databases through the
best network infrastructure (cellular or fixed); providing
end-to-end services capabilities, with three targets: the end device
(smart meter), the concentrator/gateway and the service platform.
Further, smart metering application profiles will be specified
including service functionalities.

\subsection{Worldwide standardization}

The outcome of any deployment of smart grid/metering systems will
depend on successful, quick and future-proof standardization of the
major, long-lived components of the system, including the
communications. In this section we have given an overview of smart
metering standardization activities in Europe. However, it has to be
mentioned that there are other major smart grid standards worldwide,
for example, in US notably IEEE P2030 \cite{67}, ANSI \cite{68}, US
NIST \cite{37} and future IP for smart grids in the IETF \cite{69}.
For instance, it is important to note the development of the ANSI
C12 suite of standards in the USA, that have been developed for
electricity meters, in a similar capacity to the standards under the
aegis of CLC/TC 13 in Europe.  These standards are now being updated
to reflect advances in smart metering, e.g. with the introduction
of, among others, the C12.19 standard for Utility Industry End
Device Data Tables (data models and formats for metering data) and
the C12.22 standard for Protocol Specification for Interfacing to
Data Communication Networks (communicating smart metering data
across a network) \cite{70}.

The IEEE P2030 \cite{67} project addresses smart grid
interoperability and is composed of three task force groups looking
at different aspects of interoperability in power systems (Task
Force 1), information systems (Task Force 2) and communication
systems (Task Force 3). The aim of this project is to provide
guidelines for enabling integration of energy technology and
information and communications technology (ICT) to achieve seamless
operation of the grid components and a more reliable and flexible
electric power system.

The IEEE 1547 standard was approved in October 2003 and outlines the
collection of requirements and specifications for interconnecting
distributed energy resources to the distribution segment of the
electric power system \cite{80}. The outlined requirements are
relevant to the performance, operation, testing, safety, and
maintenance of the interconnection. They are globally needed for
interconnection of distributed energy resources including both
distributed generators and energy storage infrastructure, which is
essential to realize the goals of smart grid.

In China, smart grid standardization is led by the State Grid
Corporation of China (SGCC). SGCC has recently signed a strategic
cooperation agreement with General Electric (GE) and the Chinese
Academy of Science to jointly develop smart grid standards.
Standardization activities are expected in the following technical
areas: electric vehicle charging, grid-scale energy storage
integration, distributed resources, and microgrids \cite{71}. In
India, the IEEE Standards Association (IEEE-SA) has introduced two
new standards, IEEE 1701 and IEEE 1702, to create a multi-source
plug and play communications environment for diverse smart metering
devices. Both standards provide lower layer communication protocols
for LAN/WAN ports and telephone ports, respectively, used in
conjunction with utility metering \cite{72}.  In Japan,
standardization activities have been focused on the vision of
``Smart Community'' which involves the integration of smart grid,
energy storage, electric vehicles, and intelligent transport
systems. The newly established Japan Smart Community Alliance is
coordinating industrial efforts in this area \cite{73}.

\section{Conclusion}

In this paper we have presented an overview of the unique challenges
and opportunities posed by smart grid communications, e.g.
interoperability, new infrastructure requirements, scalability,
demand response, security and privacy. The success of future smart
grid depends heavily on the communication infrastructure, devices,
and enabling services and software. Results from much existing
communications research can be potentially applied to the extremely
large-scale and complex smart grid, which will become a killer
application. In parallel to technical issues of smart grids, we have
also discussed the current status of standardization on smart
metering in Europe. It is very desirable to have a single set of
standards defining the interfaces, communications and data exchange
formats for smart metering and smart grids in Europe.  However, due
to the current pressure on deploying smart metering solutions at
different timescales in different countries, and for different
energy supply companies, the timely harmonization of the many
existing standards with the new additional functionality
requirements will be very difficult.  It is very important that the
European standardization activities are aligned and take into
account these requirements, and reflect them well at international
standardization activities.

Although the roadmap of worldwide smart grid deployment is still not
clear, it is almost certain that the future intelligent energy
network empowered by advanced ICT technology will not only be as big
as the current Internet, but also change people's lives in a
fundamental way similar to the Internet. As communication is an
underpinning technology for this huge development, we envisage that
smart grids will be an exciting research area for communication
engineers for many years to come.


%

\section*{Acknowledgment}

The authors would like to thank their colleagues at Toshiba Research
Europe for helpful discussions and its Directors for their support
of this work. The detailed comments from the anonymous reviewers are
gratefully acknowledged.



\bibliographystyle{IEEEtran}
%

\bibliography{ref2}

\begin{thebibliography}{10}
\providecommand{\url}[1]{#1}
\csname url@samestyle\endcsname
\providecommand{\newblock}{\relax}
\providecommand{\bibinfo}[2]{#2}
\providecommand{\BIBentrySTDinterwordspacing}{\spaceskip=0pt\relax}
\providecommand{\BIBentryALTinterwordstretchfactor}{4}
\providecommand{\BIBentryALTinterwordspacing}{\spaceskip=\fontdimen2\font plus
\BIBentryALTinterwordstretchfactor\fontdimen3\font minus
  \fontdimen4\font\relax}
\providecommand{\BIBforeignlanguage}[2]{{%
\expandafter\ifx\csname l@#1\endcsname\relax
\typeout{** WARNING: IEEEtran.bst: No hyphenation pattern has been}%
\typeout{** loaded for the language `#1'. Using the pattern for}%
\typeout{** the default language instead.}%
\else
\language=\csname l@#1\endcsname
\fi
#2}}
\providecommand{\BIBdecl}{\relax}
\BIBdecl

\bibitem{parc}
PARC White Paper, ``Fast demand response,'' 2010.


\bibitem{lindley}
D.~Lindley, ``The energy storage problem,'' \emph{Nature}, vol. 463, Jan. 2010.

\bibitem{faruqui}
A.~Faruqui, D.~Harris, and R.~Hledik, ``Unlocking the 53 billion euros savings
  from smart meters in the {EU}: How increasing the adoption of dynamic tariffs
  could make or break the {EU}'s smart grid investments,'' \emph{Energy
  Policy}, vol.~38, no.~10, 2010.

\bibitem{booth}
A.~Booth, M.~Greene, and H.~Tai, ``{US} smart grid value at stake: The \$ 130
  billion question,'' McKinsey Report on Smart Grid, Summer 2010.

\bibitem{netl}
National Energy Technology Laboratory for the U.S. Department of Energy, Office of Electricity 
Delivery and Energy Reliability, ``The NETL Modern Grid Initiative - Powering our 21st-Century Economy, 
Modern Grid Benefits,'' Aug. 2007.

\bibitem{doe}
US Department of Energy, ``Smart Grid Research and Development: Multi-Year Program 
Plan ({MYPP}) 2010-2014,'' Available at http://www.oe.energy.gov.

\bibitem{ibm}
O.~Sundstrom, D.~Gantenbein, and C.~Binding, ``Adding {G}ridwise information to
  {E}cogrid deliverable {D}1.1,'' IBM, Tech. Rep., 2011.

\bibitem{mexico}
New Mexico Business Weekly, ``Japan signs smart grid accords with {N}ew {M}exico,'' March 2010.

\bibitem{ec1}
European Commission, ``European smart grids technology platform - vision and strategy
  for {E}urope's electricity networks of the future,'' Directorate-General for
  Research - Sustainable Energy Systems, 2006.

\bibitem{ec2}
European Commission, ``Commission Working Document - Consultation on the future {EU} 2020 strategy,'' 2009.

\bibitem{fippp}
European Commission, ``Future {I}nternet for future {E}uropean economies and societies,'' 2010.

\bibitem{siemens}
FINSENY Consortium, {FIPPP FINSENY}, http://www.fi-ppp.eu/projects/finseny/, 2010.

\bibitem{91}
C.~Lo and N.~Ansari, ``The progressive smart grid system from both power and
  communications aspects,'' \emph{IEEE Communications Surveys and Tutorials},
  2012.

\bibitem{92}
X.~Fang, S.~Misra, G.~Xue, and D.~Yang, ``Smart grid - the new and improved
  power grid: A survey,'' \emph{IEEE Communications Surveys and Tutorials},
  2012.

\bibitem{13}
F.~Li, W.~Qiao, H.~Sun, H.~Wan, J.~Wang, Y.~Xia, Z.~Xu, and P.~Zhang, ``Smart
  transmission grid: Vision and framework,'' \emph{IEEE Trans. Smart Grid},
  vol.~1, no.~2, 2010.

\bibitem{14}
P.~Zhang, F.~Li, and N.~Bhatt, ``Next generation monitoring, analysis and
  control for the future smart control center,'' \emph{IEEE Trans. Smart Grid},
  vol.~1, no.~2, Sep. 2010.

\bibitem{15}
J.~McDonald, ``Leader or follower, developing the smart grid business case,''
  \emph{IEEE Power and Energy Magazine}, Nov. 2008.

\bibitem{16}
Z.~Fan, G.~Kalogridis, C.~Efthymiou, M.~Sooriyabandara, M.~Serizawa, and
  J.~McGeehan, ``The new frontier of communications research: Smart grid and
  smart metering,'' in \emph{ACM e-Energy}, 2010.

\bibitem{90}
T.~Khalifa, K.~Naik, and A.~Nayak, ``A survey of communication protocols for
  automatic meter reading applications,'' \emph{IEEE Communications Surveys and
  Tutorials}, vol.~13, no.~2, 2011.

\bibitem{17}
DLMS User Association, www.dlms.com, 2011.

\bibitem{18}
IEC 62056, ``Electricity metering - data exchange for meter reading, tariff and
  load control,'' \emph{International Electrotechnical Commission series of
  standards}, 2002.

\bibitem{85}
S.~Feuerhahn, M.~Zillgith, C.~Wittwer, and C.~Wietfeld, ``Comparison of the
  communication protocols {DLMS/COSEM, SML} and {IEC} 61850 for smart metering
  applications,'' in \emph{Second IEEE International Conference on Smart Grid
  Communications}, 2011.

\bibitem{81}
M.~Souryal and N.~Golmie, ``Analysis of advanced metering over a wide area
  cellular network,'' in \emph{Second IEEE International Conference on Smart
  Grid Communications}, 2011.

\bibitem{82}
G.~Iyer, P.~Agrawal, E.~Monnerie, and R.~Cardozo, ``Performance analysis of
  wireless mesh routing protocols for smart utility networks,'' in \emph{Second
  IEEE International Conference on Smart Grid Communications}, 2011.

\bibitem{19}
M/441 EN, ``Standardization mandate to {CEN, CENELEC and ETSI} in the field of
  measuring instruments for the development of an open architecture for utility
  meters involving communication protocols enabling interoperability,'' European 
  Commission - Enterprise and Industry Directorate-General, 2009.

\bibitem{20}
Smart Meters Co-ordination Group (SM-CG), ``Interim response report to {M}/441,'' Dec. 2009.

\bibitem{21}
C.~Chong and S.~Kumar, ``Sensor networks: evolution, opportunities, and
  challenges,'' \emph{Proceedings of the IEEE}, vol.~91, no.~8, 2003.

\bibitem{22}
A.~Wheeler, ``Commercial applications of wireless sensor networks using
  {Z}ig{B}ee,'' \emph{IEEE Communications Magazine}, Apr. 2007.

\bibitem{23}
G.~Wu, S.~Talwar, K.~Johnsson, N.~Himayat, and K.~Johnson, ``{M2M}: From mobile
  to embedded {I}nternet,'' \emph{IEEE Communications Magazine}, Apr. 2011.

\bibitem{24}
N.~Kushalnagar, G.~Montenegro, and C.~Schumacher, ``{IP}v6 over low-power
  wireless personal area networks (6{L}o{WPAN}s),'' \emph{RFC 4919}, 2007.

\bibitem{25}
M.~Dohler, T.~Watteyne, T.~Winter, and D.~Barthel, ``Routing requirements for
  urban low-power and lossy networks,'' \emph{RFC 5548}, 2009.

\bibitem{26}
P.~Kulkarni, S.~Gormus, Z.~Fan, and B.~Motz, ``A self-organising mesh
  networking solution based on enhanced {RPL} for smart metering
  communications,'' in \emph{IEEE WoWMoM Workshop on Hot Topics in Mesh
  Networking}, 2011.

\bibitem{89}
T.~Watteyne, A.~Molinaro, M.~Richichi, and M.~Dohler, ``From {MANET} to {IETF
  ROLL} standardization: A paradigm shift in {WSN} routing protocols,''
  \emph{IEEE Communications Surveys and Tutorials}, vol.~13, no.~4, 2011.

\bibitem{27}
TC M2M, ``{TR} 102 691 - machine-to-machine communications ({M2M}); smart
  metering use cases,'' ETSI, 2010.

\bibitem{28}
M.~Bernaschi, F.~Cacace, G.~Lannello, S.~Za, and A.~Pescape, ``Seamless
  internetworking of {WLAN}s and cellular networks: architecture and
  performance issues in a mobile {IP}v6 scenario,'' \emph{IEEE Wireless
  Communications}, vol.~12, no.~3, 2005.

\bibitem{29}
W.~Luan, D.~Sharp, and S.~Lancashire, ``Smart grid communication network
  capacity planning for power utilities,'' in \emph{IEEE PES Transmission and
  Distribution Conference and Exposition}, 2010.

\bibitem{30}
S.~Galli, A.~Scaglione, and Z.~Wang, ``Power line communications and the smart
  grid,'' in \emph{First IEEE International Conference on Smart Grid
  Communications}, 2010.

\bibitem{31}
T.~Sauter and M.~Lobashov, ``End-to-end communication architecture for smart
  grids,'' \emph{IEEE Transactions on Industrial Electronics}, Apr. 2011.

\bibitem{32}
IEEE 802.15 Smart Utility Networks (SUN) Task Group 4g, http://www.ieee802.org/15/pub/tg4g.html, 2011.

\bibitem{33}
D.~Wang, Z.~Tao, J.~Zhang, and A.~Abouzeid, ``{RPL} based routing for advanced
  metering infrastructure in smart grid,'' in \emph{IEEE ICC SG Workshop},
  2010.

\bibitem{34}
B.~Jansen, C.~Binding, O.~Sundstrom, and D.~Gantenbein, ``Architecture and
  communication of an electric vehicle virtual power plant,'' in \emph{First
  IEEE International Conference on Smart Grid Communications}, 2010.

\bibitem{35}
J.~Rosenberg, H.~Schulzrinne, G.~Camarillo, A.~Johnston, J.~Peterson,
  R.~Sparks, M.~Handley, and E.~Schooler, ``{SIP}: Session initiation
  protocol,'' \emph{RFC 3261}, 2002.

\bibitem{36}
S.~Kabisch, A.~Schmitt, M.~Winter, and J.~Heuer, ``Interconnections and
  communications of electric vehicles and smart grids,'' in \emph{1st IEEE
  International Conference on Smart Grid Communications}, 2010.

\bibitem{87}
H.~Kim, Y.~Kim, K.~Yang, and M.~Thottan, ``Cloud-based demand response for
  smart grid: Architecture and distributed algorithm,'' in \emph{Second IEEE
  International Conference on Smart Grid Communications}, 2011.

\bibitem{88}
Y.~Kim, V.~Kolesnikov, H.~Kim, and M.~Thottan, ``{SSTP}: a scalable and secure
  transport protocol for smart grid data collection,'' in \emph{Second IEEE
  International Conference on Smart Grid Communications}, 2011.

\bibitem{93}
Y.~Qian, Y.~Yan, H.~Sharif, and D.~Tipper, ``A survey on cyber security for
  smart grid communications,'' \emph{IEEE Communications Surveys and
  Tutorials}, 2012.

\bibitem{74}
X.~Hong, P.~Wang, J.~Kong, Q.~Zheng, and J.~Liu, ``Effective probabilistic
  approach protecting sensor traffic,'' in \emph{IEEE MILCOM}, 2005.

\bibitem{37}
A.~Lee and T.~Brewer, ``Guidelines for smart grid cyber security: Vol. 1, smart
  grid cyber security strategy, architecture, and high-level requirements,'' NISTIR 7628, 2010.

\bibitem{38}
AMI-SEC TF, ``{AMI} system security requirements,'' OpenSG, 2008.

\bibitem{39}
Stuxnet, http://en.wikipedia.org/wiki/stuxnet, 2010.

\bibitem{40}
R.~Anderson and S.~Fuloria, ``Who controls the off switch,'' in \emph{First
  IEEE International Conference on Smart Grid Communications}, 2010.

\bibitem{41}
W.~Stallings, \emph{Network security essentials: applications and
  standards}.\hskip 1em plus 0.5em minus 0.4em\relax Prentice Hall, 2007.

\bibitem{42}
R.~Anderson, \emph{Security Engineering}.\hskip 1em plus 0.5em minus
  0.4em\relax Wiley, 2008.

\bibitem{44}
E.~Quinn, ``Privacy and the new energy infrastructure,'' \emph{Social
  Science Research Network (SSRN)}, 2009.

\bibitem{75}
S.~Rajagopalan, L.~Sankar, S.~Mohajer, and V.~Poor, ``Smart meter privacy: A
  utility-privacy framework,'' in \emph{2nd IEEE International Conference on
  Smart Grid Communications}, 2011.

\bibitem{45}
R.~Stallman, ``Is digital inclusion a good thing? {H}ow can we make sure it
  is,'' \emph{IEEE Communications Magazine}, Feb. 2010.

\bibitem{46}
G.~Hart, ``Nonintrusive appliance load monitoring,'' \emph{Proceedings of the
  IEEE}, Dec. 1992.

\bibitem{47}
H.~Lam, G.~Fung, and W.~Lee, ``A novel method to construct taxonomy of
  electrical appliances based on load signatures,'' \emph{IEEE Transactions on
  Consumer Electronics}, May 2007.

\bibitem{48}
A.~Prudenzi, ``A neuron nets based procedure for identifying domestic
  appliances pattern-of-use from energy recordings at meter panel,'' in
  \emph{IEEE Power Engineering Society Winter Meeting}, 2002.

\bibitem{49}
Z.~Taysi, M.~Guvensan, and T.~Melodia, ``Tiny{EARS}: Spying on house appliances
  with audio sensor nodes,'' in \emph{ACM BuildSys}, 2010.

\bibitem{76}
G.~Kalogridis and S.~Denic, ``Data mining and privacy of personal behavior
  types in smart grid,'' in \emph{IEEE International Conference on Data Mining
  Workshop (ICDMW)}, 2011.

\bibitem{43}
NARUC, ``Resolution urging the adoption of general privacy principles for state
  commission use in considering the privacy implications of the use of utility
  customer information,'' Oct. 2009.

\bibitem{50}
C.~Efthymiou and G.~Kalogridis, ``Smart grid privacy via anonymization of smart
  metering data,'' in \emph{First IEEE Smart Grid Communications Conference},
  2010.

\bibitem{51}
J.~Bohli, C.~Sorge, and O.~Ugus, ``A privacy model for smart metering,'' in
  \emph{IEEE International Conference on Communications (ICC) SG Workshop},
  2010.

\bibitem{77}
F.~Garcia and B.~Jacobs, ``Privacy-friendly energy-metering via homomorphic
  encryption,'' in \emph{6th Workshop on Security and Trust Management (STM)},
  2010.

\bibitem{78}
Y.~Kim, E.~Ngai, and M.~Srivastava, ``Cooperative state estimation for
  preserving privacy of user behaviors in smart grid,'' in \emph{2nd IEEE
  International Conference on Smart Grid Communications}, 2011.

\bibitem{79}
L.~Sankar, S.~Kar, R.~Tandon, and V.~Poor, ``Competitive privacy in the smart
  grid: An information-theoretic approach,'' in \emph{2nd IEEE International
  Conference on Smart Grid Communications}, 2011.

\bibitem{53}
G.~Kalogridis, C.~Efthymiou, T.~Lewis, S.~Denic, and R.~Cepeda, ``Privacy for
  smart meters: Towards undetectable appliance load signatures,'' in
  \emph{First IEEE International Conference on Smart Grid Communications},
  2010.

\bibitem{52}
P.~McDaniel and S.~McLaughlin, ``Security and privacy challenges in the smart
  grid,'' \emph{IEEE Security and Privacy}, May 2009.

\bibitem{54}
Electricity demand, http://www.mpoweruk.com/electricity\textunderscore
  demand.htm, 2010.

\bibitem{55}
S.~Gormus, P.~Kulkarni, and Z.~Fan, ``The power of networking: How networking
  can help power management,'' in \emph{First IEEE International Conference on
  Smart Grid Communications}, 2010.

\bibitem{56}
K.~Letaief and Y.~Zhang, ``Dynamic multiuser resource allocation and adaptation
  for wireless systems,'' \emph{IEEE Wireless Communications}, Aug. 2006.

\bibitem{57}
N.~Hatziargyriou, H.~Asano, R.~Iravani, and C.~Marnay, ``Microgrids: An
  overview of ongoing research, development, and demonstration projects,''
  \emph{IEEE Power and Energy Magazine}, pp. 1488--1505, Aug. 2007.

\bibitem{58}
S.~Gormus, D.~Kaleshi, J.~McGeehan, and A.~Munro, ``Performance comparison of
  cooperative and non-cooperative relaying mechanisms in wireless networks,''
  in \emph{IEEE WCNC}, 2006.

\bibitem{59}
A.~Mohsenian-Rad, V.~Wong, J.~Jatskevich, and R.~Schober, ``Optimal and
  autonomous incentive-based energy consumption scheduling algorithm for smart
  grid,'' in \emph{IEEE ISGT}, 2010.

\bibitem{60}
Z.~Fan, ``Distributed demand response and user adaptation in smart grids,'' in
  \emph{IEEE/IFIP IM}, 2011.

\bibitem{61}
------, ``Distributed charging of {PHEV}s in a smart grid,'' in \emph{IEEE
  International Conference on Smart Grid Communications}, 2011.

\bibitem{62}
S.~Bu, F.~Yu, and P.~Liu, ``Stochastic unit commitment in smart grid
  communications,'' in \emph{IEEE INFOCOM 2011 Workshop on Green Communications
  and Networking}, 2011.

\bibitem{83}
S.~Yue, J.~Chen, Y.~Gu, C.~Wu, and Y.~Shi, ``Dual-pricing policy for
  controller-side strategies in demand side management,'' in \emph{Second IEEE
  International Conference on Smart Grid Communications}, 2011.

\bibitem{84}
P.~Samadi, R.~Schober, and V.~Wong, ``Optimal energy consumption scheduling
  using mechanism design for the future smart grid,'' in \emph{Second IEEE
  International Conference on Smart Grid Communications}, 2011.

\bibitem{86}
K.~Wang, S.~Low, and C.~Lin, ``How stochastic network calculus concepts help
  green the power grid,'' in \emph{Second IEEE International Conference on
  Smart Grid Communications}, 2011.

\bibitem{63}
National Institute of Standards and Technology, ``{NIST} framework and roadmap for smart grid
  interoperability standards, release 1.0,'' Jan. 2010.

\bibitem{64}
ZigBee Standards Organization, ``Zig{B}ee smart energy profile specification,'' Document 075356r15, Dec. 2008.

\bibitem{65}
Arch Rock, ``Smart grid standards - meter reading and control using {IEC}
  61968-9,'' http://www.archrock.com/blog/tag/iec/, 2010.

\bibitem{66}
TC M2M, ``Terms of reference ({T}o{R}) for technical committee
  machine-to-machine communications ({M}2{M}),'' ETSI, 2009.

\bibitem{67}
IEEE P2030, http://grouper.ieee.org/groups/scc21/2030/2030\textunderscore
  index.html, 2010.

\bibitem{68}
ANSI, http://webstore.ansi.org/, 2010.

\bibitem{69}
F.~Baker and D.~Meyer, ``Internet protocols for the smart grid,'' \emph{IETF
  Internet Draft}, Apr. 2011.

\bibitem{70}
A.~Snyder and M.~Stuber, ``The {ANSI C}12 protocol suite - updated and now with
  network capabilities,'' in \emph{Power Systems Conference: Advanced Metering,
  Protection, Control, Communication, and Distributed Resources}, 2007.

\bibitem{80}
IEEE 1547, http://grouper.ieee.org/groups/scc21/1547/1547\textunderscore
  index.html, 2010.

\bibitem{71}
K.~Ziegler, ``Chinese standardization in smart grids: a {E}uropean perspective,''
  http://www.talkstandards.com/chinese-standardization-in-smart-grids-a-europe%
an-perspective/, 2011.

\bibitem{72}
OpenPR, ``{IEEE} introduces advanced metering standards for the first time in
  {I}ndia,''
  http://www.openpr.com/pdf/191439/IEEE-Introduces-Advanced-Metering-Standards%
-for-the-first-time-in-India.pdf, 2011.

\bibitem{73}
H.~Nakanishi, ``Smart grid standardization activities in {J}apan,''
  http://www.ksgw.or.kr/down/pr/KSGW\textunderscore
  HironoriNakanishi(101111).pdf, 2011.

\end{thebibliography}

%

\begin{IEEEbiographynophoto}{Zhong Fan} is a Chief Research Fellow with Toshiba Research Europe in Bristol,
UK. Prior to joining Toshiba, he worked as a Research Fellow at
Cambridge University, a Lecturer at Birmingham University and a
Researcher at Marconi Labs Cambridge. He was also awarded a BT
Short-Term Fellowship to work at BT Labs. He received his BS and MS
degrees in Electronic Engineering from Tsinghua University, China
and his PhD degree in Telecommunication Networks from Durham
University, UK. His research interests are wireless networks, IP
networks, M2M, and smart grid communications.
\end{IEEEbiographynophoto}

\begin{IEEEbiographynophoto}
{Parag Kulkarni} is a Principal Research Engineer at the
Telecommunications Research Laboratory of Toshiba Research Europe
Limited, Bristol, UK. His work focuses on systems aspects related to
wired/wireless communication networks and their management,
communication/networking aspects related to energy management and
intelligent context aware adaptation algorithms. He holds a
Bachelors degree in Computer Science and Engineering from the
National Institute of Technology Calicut, India and a PhD degree
from the University of Ulster, Northern Ireland.
\end{IEEEbiographynophoto}

\begin{IEEEbiographynophoto}
{Sedat Gormus} received a B.Sc. degree in Computer Science from
Karadeniz Technical University, Trabzon, Turkey in 1999. He then
worked as research and development engineer in several Turkish and
British technology companies. In 2004, He was awarded a PhD
scholarship as part of OSIRIS project at the University of Bristol,
UK. He finished his PhD studies in Cooperative Relaying Networks in
2008 and joined the Telecommunications Research Laboratory of
Toshiba Research Europe Ltd. in March 2009 as a Senior Research
Engineer in wireless communications field. He has been involved in a
number of EU and UK Technology Strategy Board projects such as
ViewNet. His current research activities include smart grid
communications and wireless sensor networks.
\end{IEEEbiographynophoto}

\begin{IEEEbiographynophoto}{Costas Efthymiou} photograph and biography not available at the time of
publication.
\end{IEEEbiographynophoto}

\begin{IEEEbiographynophoto}
{Georgios Kalogridis} is a Senior Research Engineer at Toshiba
Telecommunications Research Laboratory in Bristol, UK. His current
research focuses on energy management, smart grid control, and smart
metering privacy protection. More generally, his research interests
lie in the areas of information security and privacy, combinatorial
designs, network reliability, wireless mesh networks and mobile
agents. He is the inventor of numerous patents granted in the UK and
worldwide, he has published papers at major journals and
conferences, and he has been actively involved in collaborative
projects and ETSI standardization activities. Georgios has received
an MSc in Advanced Computing from University of Bristol (UK) and a
Diploma in Electrical and Computer Engineering from University of
Patras (Greece). He is also due to his final examination for a PhD
in Mathematics from Royal Holloway, University of London.
\end{IEEEbiographynophoto}

\begin{IEEEbiographynophoto}{Mahesh Sooriyabandara} photograph and biography not available at the time of
publication.
\end{IEEEbiographynophoto}

\begin{IEEEbiographynophoto}
{Ziming Zhu} received the MSc degree in Communication Networks and
Signal Processing with Distinction from Bristol University, U.K., in
October 2010. He is currently a Ph.D. student within the Advanced
Signal Processing Group (ASPG) at Loughborough University, U.K.  His
research interests include resource management and optimization for
wireless networks and smart grids.
\end{IEEEbiographynophoto}

\begin{IEEEbiographynophoto}
{Sangarapillai Lambotharan} received the Ph.D. degree in signal
processing from Imperial College London, U.K., in 1997. He was with
Imperial College until 1999 as a Postdoctoral Research Associate. In
1996, he was a Visiting Scientist with the Engineering and Theory
Center, Cornell University, Ithaca, NY. From 1999 to 2002, he was
with the Motorola Applied Research Group, U.K., as a Research
Engineer, and investigated various projects, including physical-link
layer modeling and performance characterization of GPRS, EGPRS, and
UTRAN. From 2002 to 2007, he was with the King's College London,
London, U.K., and Cardiff University, Wales, U.K., as a Lecturer and
Senior Lecturer, respectively. In September 2007, he joined the
Advanced Signal Processing Group, Loughborough University,
Leicestershire, U.K., as a Reader, and was promoted to Professor of
digital communications in September 2011. He has published more than
100 conference and journal articles in these areas. He serves as an
Associate Editor for the EURASIP Journal on Wireless Communications
and Networking. His current research interests include
multiple-input-multiple-output, wireless relay networks, cognitive
radio networks, and smart grids.
\end{IEEEbiographynophoto}

\begin{IEEEbiographynophoto}{Woon Hau Chin} photograph and biography not available at the time of
publication.
\end{IEEEbiographynophoto}


\vfill


\end{document}